\documentclass[manuscript,screen]{acmart}

\usepackage{microtype}
\usepackage{graphicx}
\usepackage{subfigure}
\usepackage{booktabs} 
\usepackage{wrapfig}
\usepackage{hyperref}

\usepackage{amsmath}
\usepackage{mathtools}
\usepackage{amsthm}
\usepackage{balance}
\usepackage{enumitem}%
\colorlet{BLUE}{blue}
\colorlet{RED}{red}
\colorlet{GREEN}{green}
\colorlet{CYAN}{cyan}
\colorlet{BROWN}{brown}
\colorlet{PURPLE}{purple}
\colorlet{ORANGE}{orange}
\colorlet{TEAL}{teal}

\begin{document}

\title{Copyright Protection in Generative AI: A Technical Perspective}

\author{Jie Ren}
\email{renjie3@msu.edu}
\affiliation{%
  \institution{Michigan State University}
  \country{USA}
}
\author{Han Xu}
\affiliation{%
  \institution{Michigan State University}
  \country{USA}}
\email{xuhan1@msu.edu}

\author{Pengfei He}
\affiliation{%
  \institution{Michigan State University}
  \country{USA}}
\email{hepengf1@msu.edu}

\author{Yingqian Cui}
\affiliation{%
  \institution{Michigan State University}
  \country{USA}}
\email{cuiyingq@msu.edu}

\author{Shenglai Zeng}
\affiliation{%
  \institution{Michigan State University}
  \country{USA}}
\email{zengshe1@msu.edu}

\author{Jiankun Zhang}
\affiliation{%
  \institution{School of Artificial Intelligence, Jilin University}
  \country{China}}
\affiliation{%
  \institution{International Center of Future Science, Jilin University}
  \country{China}}
\affiliation{%
  \institution{Engineering Research Center of Knowledge-Driven Human-Machine Intelligence, MOE}
  \country{China}}
\email{zhangjk9920@mails.jlu.edu.cn}

\author{Hongzhi Wen}
\affiliation{%
  \institution{Michigan State University}
  \country{USA}}
\email{wenhongz@msu.edu}

\author{Jiayuan Ding}
\affiliation{%
  \institution{Michigan State University}
  \country{USA}}
\email{dingjia5@msu.edu}

\author{Pei Huang}
\affiliation{%
  \institution{Stanford University}
  \country{USA}}
\email{huangpei@stanford.edu}

\author{Lingjuan Lyu}
\affiliation{%
  \institution{Sony AI}
  \country{Japan}}
\email{lingjuan.lv@sony.com}

\author{Hui Liu}
\affiliation{%
  \institution{Michigan State University}
  \country{USA}}
\email{liuhui7@msu.edu}

\author{Yi Chang}
\affiliation{%
  \institution{School of Artificial Intelligence, Jilin University}
  \country{China}}
\affiliation{%
  \institution{International Center of Future Science, Jilin University}
  \country{China}}
\affiliation{%
  \institution{Engineering Research Center of Knowledge-Driven Human-Machine Intelligence, MOE}
  \country{China}}
\email{yichang@jlu.edu.cn}

\author{Jiliang Tang}
\affiliation{%
  \institution{Michigan State University}
  \country{USA}}
\email{tangjili@msu.edu}

\renewcommand{\shortauthors}{Ren et al.}

\begin{abstract}
  Generative AI has witnessed rapid advancement in recent years, expanding their capabilities to create synthesized content such as text, images, audio, and code. The high fidelity and authenticity of contents generated by these Deep Generative Models (DGMs) have sparked significant copyright concerns. There have been various legal debates on how to effectively safeguard copyrights in DGMs. This work delves into this issue by providing a comprehensive overview of copyright protection from a technical perspective. We examine from two distinct viewpoints: the copyrights pertaining to the source data held by the data owners and those of the generative models maintained by the model builders.
For data copyright, we delve into methods data owners can protect their content and DGMs can be utilized without infringing upon these rights. For model copyright, our discussion extends to strategies for preventing model theft and identifying outputs generated by specific models.
Finally, we highlight the limitations of existing techniques and identify areas that remain unexplored. Furthermore, we discuss prospective directions for the future of copyright protection, underscoring its importance for the sustainable and ethical development of Generative AI.
\end{abstract}




\maketitle

\section{Introduction}





Recently, generative AI models have been extensively developed to produce a wide range of synthesized content, including text, images, audio, and code, among others. For example, advanced image generative models, such as Diffusion Models (DMs)~\citep{ho2020denoising}, can produce highly realistic and detailed photographs and paintings. Similarly, large language models (LLMs) like ChatGPT~\citep{Achiam2023GPT4TR} can be leveraged to compose coherent and creative text articles with arbitrary genres and storylines. We refer to these advanced models as ``Deep Generative Models'' (DGMs). However, because of the remarkable fidelity and authenticity of the generated contents from DGMs,  concerns have been raised regarding the associated \textbf{copyright issues}. For example, the New York Times sued OpenAI and Microsoft for using copyrighted work for training chatGPT
\footnote{\href{https://www.nytimes.com/2023/12/27/business/media/new-york-times-open-ai-microsoft-lawsuit.html}{https://www.nytimes.com/2023/12/27/business/media/new-york-times-open-ai-microsoft-lawsuit.html}, \href{https://www.nytimes.com/2024/01/08/technology/openai-new-york-times-lawsuit.html}{https://www.nytimes.com/2024/01/08/technology/openai-new-york-times-lawsuit.html}}. Midjourney was accused to output images copied from commercial films\footnote{\href{https://spectrum.ieee.org/midjourney-copyright}{https://spectrum.ieee.org/midjourney-copyright}}. These copyright issues may pertain to various parties involved in the generation process. In specifics:
\begin{itemize}[noitemsep,topsep=0pt]
    \item[(1)] \textit{Source Data Owners.} To generate high-quality contents, DGMs require training on a large amount of data, collected from various resources such as the Internet, even without the permission of the original data owner. As demonstrated in recent studies~\citep{carlini2022quantifying, carlini2023extracting}, it is likely that both DMs and LLMs can produce contents with a high coincidence to parts of the training data samples. Besides, DGMs can also be utilized to directly edit the contents or imitate the artistic styles from source images and texts. These facts raise concerns for the data owners, as DGMs can generate data that closely resembles or replicates their original data without authorization.
    \item[(2)] \textit{DGM Users.} DGMs are also frequently utilized by DGM users and assist DGM users for creative composing. However, whether the DGMs users should receive copyright for their generated contents is still a complex and evolving legal and ethical issue. For example, in 2023\footnote{\href{https://fingfx.thomsonreuters.com/gfx/legaldocs/klpygnkyrpg/AI\%20COPYRIGHT\%20decision.pdf}{Registration of ``Zarya of the Dawn''}}, US Copyright Office refused to register a graphic novel for an artist facilitated by Midjourney\footnote{\href{https://www.midjourney.com/}{https://www.midjourney.com/}} (a popular AI image generation model). However, another giant generative-AI company, OpenAI, claims that the model users own the created data via the models from OpenAI, including the right to reprint, sell, and merchandise\footnote{\href{https://openai.com/policies/terms-of-use}{https://openai.com/policies/terms-of-use}}.
    \item[(3)] \textit{DGM Providers.} The DGM providers contribute great efforts on the training of a DGM. It includes collecting and processing the large amount of data, engineering on the training and tuning for optimized model performance. Therefore, the DGM providers also have reasons to demand the copyright of the generative contents.
\end{itemize}

It has been under active discussion among government officials, lawmakers, and the general public, about who should claim the copyright and how to protect it. For instance, in early 2023, the U.S. Copyright Office initiated a process to gather feedback on copyright-related concerns related to generative AI. This included the discussions on the scope of copyright for works created via using AI tools and the use of copyrighted materials in AI training. Additionally, legal perspectives on these matters have also been explored in various opinions~\citep{zirpoli2023generative, franceschelli2022copyright, samuelson2023generative}. However, in this article, we approach the topic from a different angle, where we provide an overview on existing computational methodologies, which have been proposed for copyright protection \textbf{from a technical perspective}. These are potential viable solutions for either AI model creator or model users, for copyright protection. In general, these computational techniques can be categorized according to the receiver of the copyright: 
\begin{itemize}[noitemsep,topsep=0pt]
    \item \textit{For source data owners}, the protection on their original works could be achieved by: (a) \textit{``Crafting unrecognizeable examples''}, which refers to the process that the data owners manipulate their data to hinder DGMs from extracting information; (b) \textit{``Watermark techniques''} can be used by the data owner to trace and distinguish whether a generated work is produced based on their original creation; (c) ``\textit{Machine Unlearning}'' means that the data owner can request a deletion of their data from the model or its output, once they identify the copyright infringement; (d) ``\textit{Dataset De-duplication}'' removes duplicated data to mitigate the memorization effect to prevent the training data from being generated; (e) ``\textit{Alignment}'' which uses a reward signal to reduce the memorization in LLMs; and (f) others including improved training and generation algorithms for better behaviors of LLMs.
    \item \textit{For DGM users} who create new works assisted by DGMs, the protected object is the generated contents from DGMs. Thus, the techniques for this type of copyright is barely related to the generation process of DGMs, as traditional copyright protection strategies can be also applied for protection of the DGM generated contents.
    \item For DGM providers, there are representative ``watermarking strategies'' to inject the watermarks into the generated content or model parameters, such that we can track the ownership of the model.
\end{itemize}

Given the diversity in protection objectives, as well as DGM applications, we are motivated to have an overview on existing computational methods in this direction. Essentially, in Section~\ref{sec:image}, we will majorly discuss the copyright protection techniques for DGMs in the image domain. In Section~\ref{sec:llm}, we discuss the strategies for text generation. Finally, we discuss the related problems in other domains, such as graphs, codes and audio generation in Section~\ref{sec:others}. In each section, we will introduce the background knowledge of existing DGMs, as well as the existing methodologies for data protection under different scenarios.

\section{Copyright in Image Generation}\label{sec:image}

In this section, we first introduce background knowledge on existing popular image generation models. Then, we define the problems related to the copyright issues for these image generation models, and introduce different strategies which can be utilized for data and model copyright protection.

\subsection{Background: DGMs for Image Generation}\label{sec:DGM}

In the era of deep learning, there are various types of image generation models that have been extensively studied. Most of them follow a pipeline to first collect a set of $n$ training images $\mathcal{X} = \{x_i\}^n_{i=1}$, whose samples' distribution can be denoted as $p(x)$, and a model is trained to explicitly or implicitly estimate this data distribution. We denote the learned distribution as $p'(x)$. During the image generation stage, the new images are generated by sampling from the learned distribution $p'(x)$. For example, some popular models~\citep{goodfellow2014generative, ho2020denoising} can be written as the form of $f(\cdot)$, which takes random noise $z$ (following distributions $p_z$ such as Gaussian Distribution) as input and generates samples $f(z)$.  Besides, there are more advanced generative models, which can generate new samples based on user's particular demands. For example, there are conditional generative models~\citep{mirza2014conditional} to generate samples $f(z, c)$ which belong to a specific sub-distribution of $p(x)$, where $c$ can denote the given condition. In text-to-image models~\citep{rombach2022high}, the condition $c$ can also be formatted as a language prompt for the desired generated samples. Furthermore, there are also advanced models which can directly take inputs from a few source samples $x$, and then edit or modify them to obtain new samples $f(x)$. In the following, we provide a brief overview on the mechanisms of several popular image generative models.

\begin{itemize}
\item \textbf{Autoencoders}~\citep{vincent2010stacked, kingma2013auto} refer to the generative models, which consist of an encoder module and a decoder module for image reconstruction. In detail, the encoder projects a real image $x$ into a latent vector $z$ in the latent space. Then, the decoder takes the input of the latent vectors $z$ to generate new samples $x'$. The decoder is trained to reconstruct the original sample $x$ from the its latent $z$ produced by the encoder. In the work~\citep{kingma2013auto}, the latent vectors are further regulated to follow a standard Gaussian distribution. During the generation process, new latent vectors are sampled from the regulated distribution directly without the encoder and then decoded to obtain the generated images. 

\item \textbf{Generative Adversarial Networks (GAN)}~\citep{goodfellow2014generative} are proposed to synthesize images via solving a min-max game. In GANs, there are two adversarial players, a generator $G(\cdot)$ and a discriminator $D(\cdot)$. 
The generator aims to generate images that are highly realistic, while the discriminator aims to determining whether a sample is real or synthesized by $G(\cdot)$. By solving the min-max problem, $G(\cdot)$ and $D(\cdot)$ will eventually converge to a point at which $D(\cdot)$ is well trained to distinguish the real and fake images, but $G(\cdot)$ can generate convincing images that are still possible to confuse $D(\cdot)$. The adversarial min-max problem can be formulated as: 
\begin{equation}\label{eq: gan}
    \begin{aligned}
    \min _G \max _D \mathbb{E}_{ {x} \sim p( {x}) } \log D( {x}) +\mathbb{E}_{ {z} \sim p_{ {z}}} \log (1-D(G( {z})))
    \end{aligned}
\end{equation}
where $G(z)$ is the generated images, given the latent vector $z \sim p_z$ ($p_z$ is the predefined distribution like an isotropic Gaussian noise). In Eq.~\eqref{eq: gan} the goal of $G(\cdot)$ is to be trained to generate more authentic and realistic samples to mislead $D(\cdot)$.

\item \textbf{Diffusion Models}, like DDPM \citep{ho2020denoising}, 
are a type of generative models which generate images via two $T$-step Markov processes: a forward process and a reverse process. The forward process gradually adds noise to transform an input image $x_0$ into an isotropic Gaussian sample $x_T$. In the reverse process, a denoising neural network is trained to transform a sample from the Gaussian distribution into the image distribution. The forward process can be formulated as:
\begin{equation*}
 {x}_t = \sqrt{1-\beta_t} {x}_{t-1} + \beta_t\epsilon, 
\end{equation*}
where $\epsilon\sim \mathcal{N}(0,I)$ and $\beta_t$ controls the variance of injected noise. Then, the reverse process learns to denoise from noisy variable $x_{t+1}$ to less-noisy variable $x_t$. The denoising step is approximately equivalent to estimating the injected noise $\epsilon_t$ with a parametric neural network
$\epsilon_\theta(x_{t+1}, t)$ in practice. The network is trained to minimize
$l_2$ distance between estimated noise and true noise:
\begin{equation}\label{eq: ddpm_loss}
    \begin{aligned}
    \mathcal{L}_\text{dm} = \mathbb{E}_{t, {x}_0, {\epsilon}\sim \mathcal{N}(0,I) }\left [\left \|  {\epsilon}- {\epsilon}_{\theta} (x_{t+1}, t)) \right \|_2^{2}\right ] 
    \end{aligned}.
\end{equation}
Once the training of denoising network ${\epsilon}_{\theta}$ is complete, a new image can be generated starting from a random noise by solely employing the reverse process. 

Based on DDPM, Latent Diffusion Model (LDM) \citep{rombach2022high} is a specific class of diffusion models, which apply the forward process in the latent
space instead of the pixel (image) space. In detail, to train an LDM, an input image $x_0$ is first mapped to its latent representation $ {z}_0=\mathcal{E} ( {x}_0)$, where $\mathcal{E}(\cdot)$ is a given image encoder. Then the forward process continues by repeatedly adding noise to the latent representation $z_0$ for $T$ steps, while the reverse process generates data by the denoising network $\epsilon_\theta$.
Once the latent representation ${z}^\prime$ is generated by the reverse process, it is then decoded by a decoder model $\mathcal{D}(\cdot)$ to get the final generated image ${x}^\prime$. 

\item \textbf{Text-to-image Diffusion Models}, such as Stable Diffusion (SD)~\citep{rombach2022high}, MidJourney, and DALL$\cdot$E2, allow the model users to generate images based on language descriptions. Stable Diffusion is achieved by combining LDMs with CLIP~\citep{radford2021learning}, a powerful model that learns the connection between the concepts of human language and images. Briefly speaking, given one of a few sentences as a prompt, a new image is produced following the basic pipeline of LDM, but conditioning on the (embedding of) language information of the given prompt. As a result, the generated contents can contain the desired semantic features and patterns by the prompt. 

Recently there are more advanced techniques such as Textual Inversion~\citep{gal2022image} and DreamBooth~\citep{ruiz2023dreambooth} which further upgrade text-to-image diffusion models for customized image editing and modifying. For example, one can fine-tune SD on a few ``input images'' via the algorithm of DreamBooth, to make the model learn new objects from the input images, and then generate new images for the targeted object in different scenes.

\end{itemize}

\subsection{Copyright Issues in Image Generation}

The development of Deep Generative Models (DGMs) marks a noteworthy advancement in image generation. Nevertheless, the impressive quality and authenticity of the generated images, as well as the efficiency in producing new ones, give rise to legitimate concerns regarding copyright matters within the realm of DGMs.

\textbf{Data copyright protection}. For the source data owner, which refers to the party or individual who owns the originality of image works, their data can be intentionally or unintentionally \textbf{collected by model trainers as training samples} to construct DGMs as introduced above. For example, recent studies~\citep{carlini2023extracting, vyas2023provable} have demonstrated that popular DGMs are highly possible to completely replicate their training data samples, which is called memorization. The possibility of data replication may severely offend the ownership of the original data samples. Moreover, the development of fine-tuning strategies such as DreamBooth can greatly improve the efficiency for unauthorized parties to directly \textbf{edit or modify} the source data to obtain new samples, which also severely infringes the copyright of the original works.

\textbf{Model copyright protection}. To obtain DGMs with advanced generation performance, it is always necessary for the model trainers to invest a significant amount of funds and labor. It grants them intellectual property rights over the trained model. However, recent works also identify the possibility to steal others models~\citep{tramer2016stealing}.

\subsection{Data Copyright Protection}

In this subsection, we review the techniques for the source data owners to protect their data copyrights.
In general, we can categorize these techniques into four major types. 
\begin{itemize}[noitemsep,topsep=0pt]
    \item {\textbf{Unrecognizable Examples}}, which aim to prevent models from learning important features of protected images. This often results in the generation of either low-quality images or completely incorrect ones;
    \item {\textbf{Watermark}}, which involves inserting unnoticeable watermarks into protected images. Data owners can detect these watermarks if their data is used for training;
    \item {\textbf{Machine Unlearning}}, which aims to ablate the contribution of copyrighted data on the DGM to prevent the model from generating based on the information of the protected images;
    \item {\textbf{Dataset De-duplication}}, which mitigates the memorization by removing the duplicated data from training set.
\end{itemize}
In the four categories, unrecognizable examples and watermarks are employed from the side of source data owners. They modify their data before releasing to the public to protect the copyright. As for machine unlearning and dataset de-duplication, they are proposed for the model builders who aim to provide DGM services legally without infringement on the copyrighted data.
Next, we will introduce these strategies in detail.

\subsubsection{Unrecognizable Examples}

From the perspective of the source data owner, one major strategy for data protection is to make their data  ``unrecognizable'' to potential DGMs. In general, a data owner can consider injecting imperceptible perturbations into the protected images, such that the DGMs cannot effectively exploit useful features from them, and hence hardly generate qualified new samples. These works of ``unrecognizable examples'' can be categorized based on how the DGM is utilized to extract information from the source image: (1)~\textit{``inference-stage protection''} counteracts DGMs which operate without the need for fine-tuning on the source image $x$. These models are capable of extracting image features directly from the source image $x$ in an inference mode; (2)~\textit{``training-stage protection''} is against DGMs that is fine-tuned on the source images to extract the desired information and generate new samples based on the extracted information. Targeting on these two types of image editing manners, a variety of data copyright protection strategies are devised.

\textbf{1) Inference stage protection.} Well-trained DGMs, such as GAN models and diffusion models, can be directly used for various image generation tasks, including image-to-image synthesis and image editing. Thus, it is crucial to address {inference-stage protection} which aims at preventing models from extracting important information.

\textit{GAN-based methods}. UnGANable~\citep{li2022unganable} is the first defense system against a method commonly used for altering photographs or artistic creations, called GAN Inversion~\cite{zhu2016generative, zhu2020domain, abdal2019image2stylegan}. 
It uses adversarial examples~\citep{goodfellow2014explaining} to mislead GAN in latent space.
\citet{ruiz2020disrupting} aimed at data protection against Image-Translation GAN~\citep{liu2017unsupervised}, which is a variant of conditional GAN that directly generates new images by inputting a source image $x$ instead of a random latent vector. It can generate an image which is manipulated from the source image $x$.
They use adversarial examples to maximize the distortion in the generated image.
\citet{yeh2020disrupting} aimed to protect users' images from DeepNude \citep{Deepnude19Vox}, a deep generative software based on the image-to-image translation algorithm. This method also borrows the idea of adversarial examples. It defines adversarial loss for Nullifying Attack which maximizes the distance of features extracted by the generator $G(\cdot)$ between perturbed images and the original image, as well as adversarial loss for Distorting Attack.
Besides, \citet{huang2021initiative} expanded adversarial examples to more practical ``grey-box'' and ``black-box'' settings, which refer to the case that the data owner is unaware of the specific model that might be used for potential copyright infringement. This method adopts a surrogate model to approximate the manipulation model and update surrogate parameters and adversarial examples in an alternative manner. These examples demonstrate that the framework to generate adversarial examples can be adapted to different image manipulation models by properly designing the loss function for different specific tasks. 

\begin{wrapfigure}{r}{0.6\textwidth}
    \centering
\includegraphics[width=0.57\textwidth]{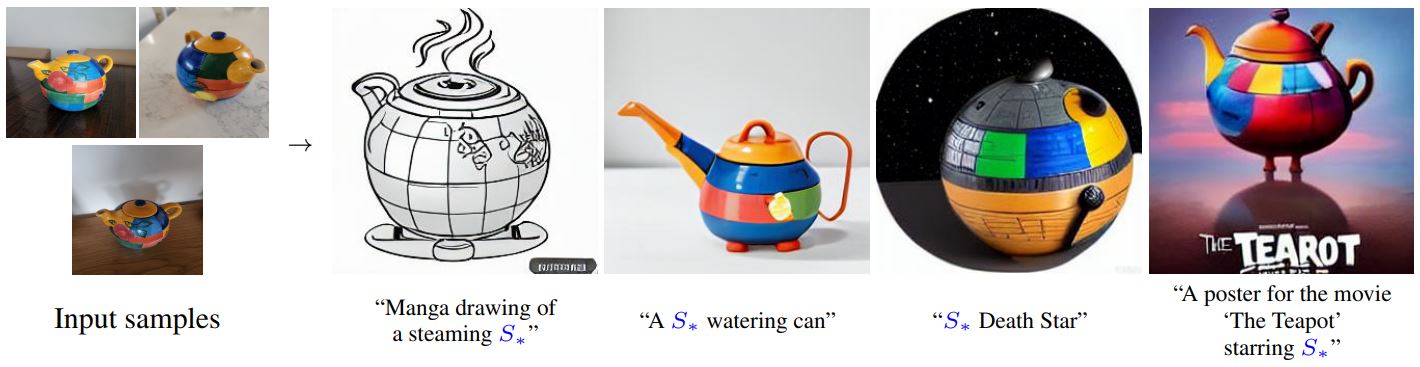}
    \vspace{-0.15in}
    \caption{\small An overview of Textual Inversion. (~\citet{gal2022image})}
    \vspace{-0.2in}
    \label{fig:textual_inversion}
\end{wrapfigure}

\textit{Diffusion models.} Beyond GAN-based models, recently diffusion models have also been exploited for various types of modifying or editing tasks, which exposes a significant risk for copyright infringement. For example, Textual Inversion~\citep{gal2022image} is an image modifying technique by Stable Diffusion, without any training or fine-tuning process. Refer to the illustration in Figure~\ref{fig:textual_inversion}, given a few source samples $x$, Textual Inversion aims to extract the knowledge from $x$ by linking the sample $x$ to a specific text string such as $S_*$. This process is achieved by adjusting the text embedding of $S_*$ in Stable Diffusion to embed the information of $x$ into it. The model user can utilize $S_*$ to compose new prompts, 
and generate new images with the information embedded in $S_*$, such as the original object or style from the source 
image, which might infringe the copyright of the source images. 
Targeted on Textual Inversion, the work ~\citep{liang2023adversarial} aims to find an adversarial example $x^{adv}$ to protect the source images $x$. $x^{adv}$ lies out of the distribution of the generated data samples of the diffusion models. Thus, the inversion process cannot find proper language tokens $S_*$ corresponding to the adversarial image.
In detail, following the general definition of the diffusion loss in Eq.~\eqref{eq: ddpm_loss}, we define the loss function on a specific sample $x$:
$\mathcal{L}_\text{dm}(x) = \mathbb{E}_{t, {\epsilon}\sim \mathcal{N}(0,I) }\left [\left \|  {\epsilon}- {\epsilon}_{\theta} (x_{t+1}, t)) \right \|_2^{2}\right ]$, the work~\citep{liang2023adversarial} aims to find a perturbation $\delta$ to maximize the loss value of the diffusion model on $x+\delta$:
\begin{align*}
\max_{ {\delta}}\mathcal{L}_\text{dm}(x +\delta)~~\text{s.t.}~~||\delta||\leq \sigma
\end{align*}
Because the diffusion model has a maximized loss on the perturbed image $x+\delta$, the sample $x+\delta$ can be seen as a natural outlier from the distribution of the generated samples of the model. As a result, the Textual Inversion process cannot effectively find reasonable token $S_*$, and thus protect the original information.

Similar to the idea of~\citep{liang2023adversarial}, the work \citep{salman2023raising} directly leverages the adversarial examples for data protection against a image editing model based on LDM. In the work of~\citep{salman2023raising}, two attack strategies are proposed:
\textit{(a) Encoder Attack}: Considering a Latent Diffusion Model is employed for image editing, given the encoder model $\mathcal{E}(\cdot)$ which maps a source image to the representation $\mathcal{E}(x)$, the encoder attack searches for an perturbation $\delta$ satisfying
$
    \min\limits_{\|\delta\|_{\infty}\le \epsilon}\|\mathcal{E}(x+\delta)-z_{target}\|^2_2,
$
where $z_{target}$ is a target latent representation. In this way, the latent representation of $x+\delta$ is close to the target representation which is pre-specified to be distinct from the original latent representation $\mathcal{E}(x)$, which severely disrupts the LDM process.
\textit{(b) Diffusion Attack}: The full image editing process can be simply denoted as a model $f(\cdot)$, and the generated image $f(x)$ can fulfill the generation goal of the model user, the diffusion attack directly finds a perturbation $\delta$ to maximize the discrepancy of the generated sample $f(x+\delta)$ and $f(x)$:
\begin{align*}
    \min\limits_{\|\delta\|_{\infty}\le \epsilon}\|f(x+\delta)-x_{target}\|^2_2
\end{align*}
As a result, the newly generated sample $f(x+\delta)$ is close to the target image $x_{target}$ thus very different from $f(x)$, which breaks the original image editing goal.
However, addressing this optimization problem is computationally costly due to the extensive number of parameters involved and the multiple steps during the diffusion process. Consequently, they suggested calculating the loss based on only a limited number of steps instead of the entire diffusion process. 

According to the discussion in~\citep{salman2023raising}, currently existing data protection strategies based on adversarial examples still face major drawbacks that can hinder their feasibility and reliability in practice:
\begin{itemize}[noitemsep,topsep=0pt]
\item \textbf{Lack of robustness to transformations.} The protected images may also be subject to image
transformations and noise purification techniques, such as cropping the image, adding filters to it, or applying a rotation. However, the authors also mention that this problem can be addressed by leveraging ``robust'' adversarial perturbations, as discussed in~\citep{athalye2018synthesizing, kurakin2018adversarial}.
\item \textbf{Generalization on different models.} The protection techniques that are designed for one generative model may not be guaranteed to be effective against future versions of these models or other types of generative models. The authors mentioned that one could hope to improve on the ``transferability' of adversarial perturbations~\citep{ren2022transferable, he2023sharpness, li2024neural}. However, these ``transferable'' perturbations will not always be transferable for all circumstances.
\end{itemize}

\textbf{2) Training stage protection} Different from directly employing DGMs to generate new images, DGMs are also usually trained or fine-tuned on some source images $x$ to effectively learn useful information from $x$ for future generations. Therefore, ``training-stage protection'' aims to add imperceptible noise to the copyrighted images, to break the training process of potential DGMs for data copyright protection. 

\begin{figure}
    \centering
\includegraphics[width=0.4\linewidth]{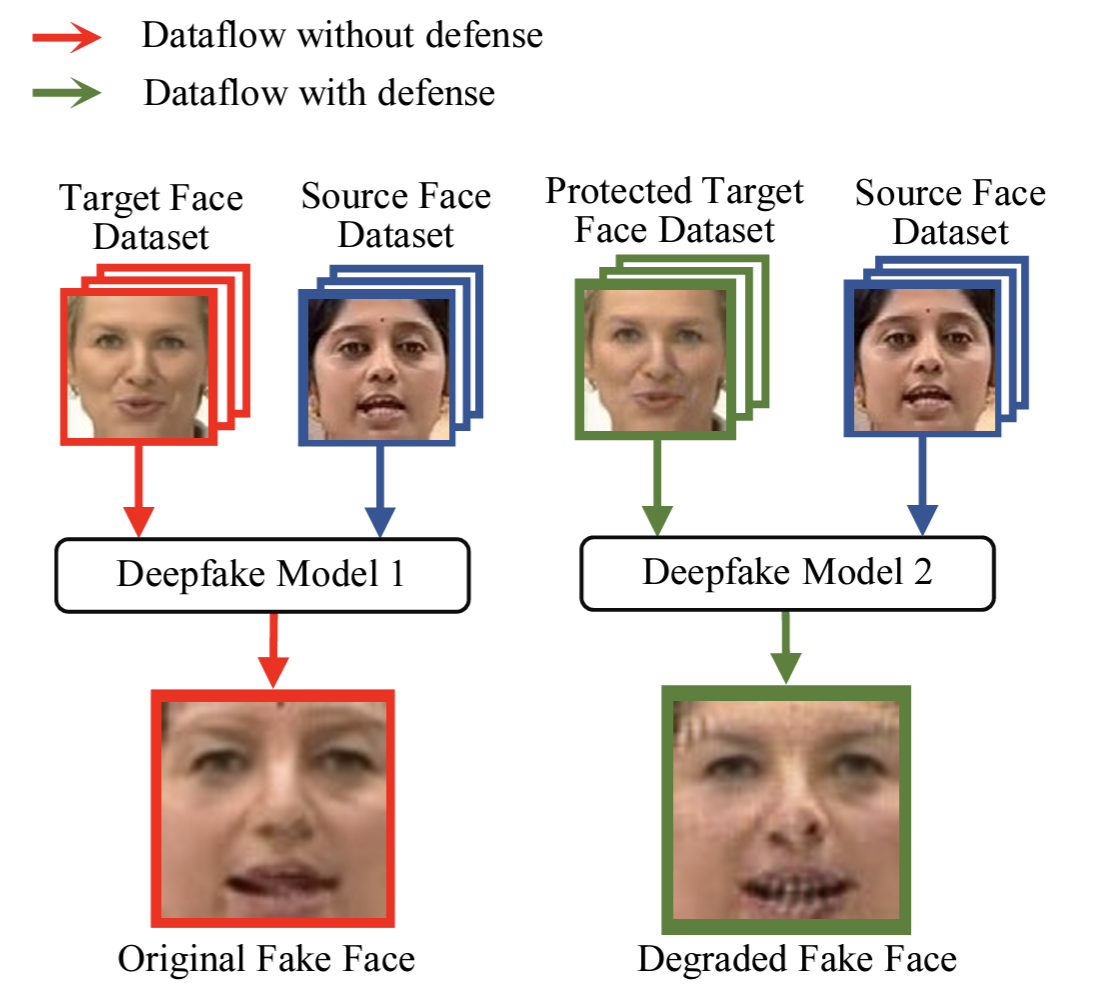}
    \caption{Transformation-aware image protection against Deepfake. (\citet{yang2021defending})}
    \label{fig:transformation-aware}
    \vspace{0.1in}
\end{figure}
This type of method is first explored on \textbf{GAN-based models}. For example, GAN-based Deepfake models~\citep{faceswap-gan2222} are representative tools that can be leveraged to swap the faces from the source images to the faces of a  target person, which
severely abuses the copyrights of both source image holders and target people. Similar tools like EditGAN \citep{ling2021editgan} and Introspective Adversarial Network \citep{brock2016neural} are developed to edit images that pose more threats to the copyrights of creative and artistic works. To handle this problem, \citet{yang2021defending} proposed to utilize the idea of adversarial examples~\citep{goodfellow2014explaining}, to break the balance in the min-max game in GAN-based DeepFake models. Specifically, they focus on Deepfake models which are trained on the target person's face images $x$ to generate other face images belonging to the target person. 
To protect the targeted face $x$ being exploited by Deepfake, they directly adopt the fast gradient sign method (FGSM) method to generate adversarial examples for GAN-based models, 
\begin{equation}\label{eq:transformation-aware}
    \begin{aligned}
 {x}^\text{adv} =  {x}+\alpha \text{sign}(\nabla_{x}\mathcal{L}_D(\text{Tr}( {x}), y_\text{real})),
\end{aligned}
\end{equation}
where $\mathcal{L}_D$ refers to the loss of the discriminator in GAN. During the generation process, the discriminator $D(\cdot)$ will 
have a large loss value on the target samples $ {x}_\text{target}$, which consequently breaks the balance of the min-max game during the training of GAN. As a result, the generated images based on the protected target images will have a degraded quality (see Figure.~\ref{fig:transformation-aware}). Notably, in Eq.\eqref{eq:transformation-aware}, a transformation operator $\text{Tr}(\cdot)$ is introduced to improve the robustness of the perturbation under various image transformations, including resizing, affine transformation and image remapping.

\citet{wang2022anti} pointed out that previous adversarial perturbations via gradient-based strategies (e.g, FGSM) could be easily removed or destroyed by image reconstructions such as MagDR~\citep{chen2021magdr} and proposed Anti-Forgery targeted on GAN-based Deepfake attacks. Anti-Forgery generates perturbations that are robust to input transformation, natural to human eyes, and applicable to black-box settings. They observed that adversarial perturbations on the LAB color space are robust to input reconstruction. Therefore, they converted the input from RGB space to the LAB color space and added perceptual-aware adversarial perturbations to the color channel to maintain robustness against input transformations including image reconstruction~\citep{chen2021magdr} and image compression~\citep{dziugaite2016study}.

\begin{wrapfigure}{r}{0.35\textwidth}  
\centering
\vspace{-0.1in}
\includegraphics[width=0.35\textwidth]{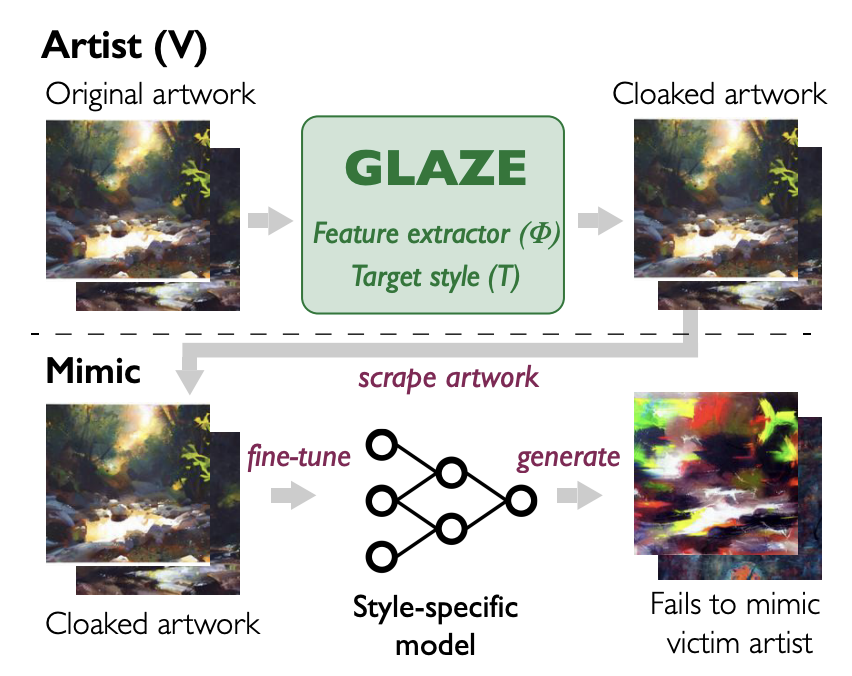}
    \vspace{-0.3in}
    \caption{\small A overview of GLAZE. (\citet{shan2023glaze})} 
    \label{fig:glaze}
    \vspace{-0.1in}
\end{wrapfigure}

Regarding \textbf{diffusion models}~\citep{rombach2022high, ruiz2023dreambooth}, which can be easily deployed to mimic the style of specific artists, via advanced fine-tuning techniques, the special structure of diffusion models (see Section~\ref{sec:DGM}) poses unique challenges, especially the denoising procedure during the reverse process can eliminate noises added to the original image~\citep{nie2022diffusion}.
GLAZE~\citep{shan2023glaze} is a representative copyright protection method focusing on text-to-image LDM and aiming at protecting artists from style mimicry. As shown in Figure~\ref{fig:glaze}, the core idea of GLAZE is to guide the diffusion model to learn an alternative target style $S_T$ that is totally different from the style of protected images. In detail, the method consists of three steps: target style choosing, style transfer, and cloak perturbation computation. GLAZE first chooses a target style $S_T$ which is sufficiently different from the protected style. A pre-trained style-transfer model $\Omega$ is utilized to transfer the protected artworks $x$ into the target style $\Omega(x,S_T)$ for optimization. Then, GLAZE computes the cloak perturbation $\delta_x$ by: 
\begin{equation}\label{eq:glaze}
\min\limits_{ {\delta}}\|\mathcal{E}(x+\delta)-\mathcal{E}(\Omega(x,S_T))\|^2_2 +\alpha \max(\text{LPIPS}( {\delta}_x)-p,0),
\end{equation}
where $\mathcal{E}(\cdot)$ is the feature extractor of the LDM.
This objective minimizes the distance between features of perturbed 
images and target-style transferred images, and LPIPS \citep{zhang2018unreasonable} constraints the perturbation to be imperceptible. When fine-tuned on protected arts, the generated images will mimic the target style rather than the arts' true style.

MIST~\citep{liang2023mist} emphasizes that existing methods generate perturbations relying on some strong assumptions on a specific model which are hard to generalize to other scenarios. For example, perturbations generated for image-to-image DGMs usually fail for Textual Inversion~\citep{salman2023raising}.
Therefore, they propose to generate perturbations that can work for various DGMs simultaneously, including DreamBooth (training-stage)~\citep{ruiz2023dreambooth}, Textual Inversion (inference-stage)~\citep{gal2022image} and image-to-image generation~\citep{rombach2022high}. To achieve this goal, they combine semantic loss from \citep{liang2023adversarial} and textual loss from \citep{salman2023raising}. They empirically show that the maximization of semantic loss leads to chaotic contents in the generated image, and the maximization of textual loss leads to a mimic of the pre-specified target image. Their empirical results reveal that perturbations from the combination of two losses can protect images under different scenarios.

Anti-DreamBooth~\citep{van2023anti} specifically targets a powerful finetuning technique,  DreamBooth~\citep{ruiz2023dreambooth}, which is proposed to personalize text-to-image diffusion models onto given source images. DreamBooth has a similar target as Textual Inversion but requires fine-tuning of diffusion models. To avoid malicious usage of DreamBooth on users' owned images, Anti-DreamBooth aims to attack the training process of DreamBooth, following the idea of data poisoning attacks. In detail, it formulates the protection problem as a bi-level optimization problem, to find perturbation $\delta$ satisfying:
\begin{align*}
        \delta^* = \max\limits_{\delta}\mathcal{L}_\text{cond}( {\theta^*},  {x}),
        ~\text{s.t.}~\theta^* =\arg\min\limits_{\theta}\mathcal{L}_\text{db}(\theta, {x}+ {\delta})~\text{and}~\| {\delta}\|\le \sigma
\end{align*}
where $\mathcal{L}_\text{db}$ denotes the training loss for DreamBooth, and $\mathcal{L}_\text{cond}(\theta^*, x)$ refers to the conditional loss of sample $x$ in prompt-guided diffusion models. By solving this problem, it searches for the perturbation, such that the fine-tuned diffusion models will disconnect the image $x$ with its corresponding language concept because of the high conditional loss. 
In this way, during the fine-tuning process, DreamBooth will overfit the adversarial images and experience worse performance in synthesizing images with high quality. Later, the work ADAF~\citep{wu2023towards} also focuses on text-to-image models but pays more attention to the text part. It points out two drawbacks of existing methods: existing methods ignore the combination of the text encoder and the image encoder; existing methods are not robust to the perturbation of prompts. Consequently, ADAF implements multi-level text-related augmentations to enhance defense stability.

\subsubsection{Watermarks}
\label{sec:image_watermark_source_data_owner}

Another approach for protection on source data copyright is to track or detect whether a suspect piece of artwork is generated by a model trained on the copyrighted data. Various AI-generated image detection methods~\citep{epstein2023online, dogoulis2023improving} can be applied to distinguish whether a sample is generated by certain models, which partially fulfill the objective. However, these methods are still not applicable to identify the source of the generated contents. Therefore, the ``watermarking'' strategy is alternatively studied. This technique involves encoding sophisticated ``identifiable information'' into the copyrighted source data, such that this information also exists in the generated samples which are trained on the watermarked images. Subsequently, a detector is leveraged to assess whether a suspect image contains this encoded information, to trace and verify the ownership of copyright.

Before DGMs, there exist various watermarking methods\citep{baluja2017hiding, hayes2017generating, vukotic2018deep, zhu2018hidden, zhang2019invisible, tancik2020stegastamp, luo2020distortion} hiding data like a message or even an image behind imperceptible perturbations. These techniques primarily concentrate on hiding information in specific images, without being specifically applied to DGMs. But the objective of protection copyright against malicious DGMs is to identify hidden messages within generated images.  Focusing on DDPM~\citep{nichol2021improved}, \citet{cui2023diffusionshield} evaluated whether the injected watermarks via previous methods for traditional image watermarks~\citep{navas2008dwt, zhu2018hidden, yu2021artificial} can still be preserved in the generated samples. The empirical results show that these methods are either partially preserved in generated images or requires large perturbation budgets. Therefore, they proposed DiffusionShield~\citep{cui2023diffusionshield}, a watermarking method designed for diffusion models.
To elaborate, blockwise watermarks, are engineered to convey a greater amount of information, allowing distinct copyright information to be more readily decoded. Then, a joint optimization strategy is leveraged to optimize both the pixel values of watermark patches, as well as a decoding model, which is utilized to detect and decode the encoded information from the generated images.

Fine-tuning text-to-image diffusion models, like Stable Diffusion, demonstrates significant potential in personalizing image synthesis and editing. Consequently, watermarking techniques are increasingly applied as a means of copyright protection during the fine-tuning phase. GenWatermark \citep{ma2023generative} is the first to propose a novel watermark system that is based on the joint learning of a watermark generator and a detector. In particular, it adopts a GAN-like structure, where a GAN generator $G$ serves as the watermark generator and a detector $D$ is trained to distinguish between clean and watermarked images. \citet{wang2023detect} aimed at a method that is independent of the choice of text-to-image diffusion models so that the perturbation can effectively protect the images from various models. In detail, they add specific stealthy transformations on the protected images as well as injecting a corresponding trigger into the caption of those images. Since they use the image warping function as the watermark generator, this method can work without a surrogate model and thus can work on different diffusion models. \citet{cui2023ft} considered a practical scenario where protectors can not control the fine-tuning process and emphasize that previous methods require many fine-tuning steps to learn the embedded watermarks. In order to make the watermark easily recognized by the model, they proposed FT-Shield which adds imperceptible perturbations that can be learned prior to the original image features, like styles and objects, by the text-to-image diffusion model. In the detection stage, a binary classifier is trained to distinguish the watermarked images and clean images. In particular, the perturbations minimize the loss of a diffusion model trained on these perturbed samples as shown in the training objective:
\begin{equation}\label{eq:ft}
    \begin{aligned}
\min\limits_{\delta}\min\limits_{\theta_1}\mathcal{L}_{dm}([\theta_1,\theta_2], x+\delta, c)
    \end{aligned}
\end{equation}
where $\theta_1$ denotes the parameters of the UNet~\citep{ronneberger2015u} which is the denoise network within the text-to-image model structure, while $\theta_2$ denotes parameters of the other parts; $x$ {and} $c$ denote the protected image and the corresponding caption, respectively. In other words, perturbation in Eq.~\eqref{eq:ft} leads to a rapid decrease in the training loss and thus serves as a 'shortcut' feature that can be quickly learned and emphasized by the diffusion model.



\subsubsection{Machine Unlearning}

In addition to the data owners, the protection of the data copyright is also considered by model builders to ensure the legal provision of DGM services. Companies try to filter out the copyrighted data from the training data. For example, Stability AI cooperated with an AI startup Spawning to build tools for data owners to claim their copyright and remove the data from the training set of Stable Diffusion, which has removed 80 million images from the training data of Stable Diffusion 3~\footnote{\href{https://the-decoder.com/artists-remove-80-million-images-from-stable-diffusion-3-training-data/}{https://the-decoder.com/artists-remove-80-million-images-from-stable-diffusion-3-training-data/}}. OpenAI also provides solution to the data owners to report the violation of data copyright~\footnote{\href{https://adguard.com/en/blog/ai-personal-data-privacy.html}{https://adguard.com/en/blog/ai-personal-data-privacy.html}}. Besides the data filtering, the model builder also takes other strategies like machine unlearning and dataset de-duplication (Section~\ref{sec:data_duplication}). 
Depending on whether the intent behind their implementation is motivated by the model builders' need to legitimize the generation process or the source data owners' requirements for the model builders, we attribute to machine unlearning~\citep{bourtoule2021machine, nguyen2022survey, zhang2023forget, kumari2023ablating} as the passive method and dataset de-duplication~\citep{webster2023duplication, somepalli2023understanding} as the active method. For the passive method, after the DGM is trained, the model builder provides an interface for data owners to claim their copyright and ablate the influence of copyrighted data from the DGM. The passive method is executed when the source data owners request. In contrast, for the active method, the model builder considers the copyright in the stage of model training. The active method is usually implemented by the model builder without the request of source data owners. In this subsection, we focus on the discussion of machine unlearning in data copyright protection.

Especially, ``Machine Unlearning''~\citep{bourtoule2021machine} refers to the protocol to make a trained model forget a specific subset of training data, by editing the model parameter to follow the distribution identical to that of a model trained without the forgotten subset. With different subsets of undesirable concepts, machine unlearning can achieve not only copyright protection, but also preserving privacy against membership inference attack and the removal of biased, NSFW and harmful concepts.  

Refer to Definition \textbf{III.1} in~\citep{bourtoule2021machine} and Section 3.1 in~\citep{nguyen2022survey}, assuming that the collected training dataset is $\mathcal{X}$, we notate the model obtained by training with the vanilla learning algorithm $\mathcal{A}$ as $\mathcal{A}(\mathcal{X})$.
Assuming the undesirable (copyrighted) subset is $\mathcal{X}_{u}$, and the unlearning mechanism is $\mathcal{U}$, the model obtained by $\mathcal{A}$ with unlearning of $\mathcal{X}_{u}$ can be represented as $\mathcal{A}_\mathcal{U}(\mathcal{X}, \mathcal{X}_{u})$. The perfect unlearning should have the distribution of parameters of unlearned model, $\mathcal{A}_\mathcal{U}(\mathcal{X}, \mathcal{X}_{u})$, to be identical to the distribution of model parameter trained on the dataset stripped of the undesirable $\mathcal{X}_{u}$, i.e.,
\begin{align}
    \mathbb{D}(\mathcal{A}_\mathcal{U}(\mathcal{X}, \mathcal{X}_{u})) = \mathbb{D}(\mathcal{A}(\mathcal{X} \backslash \mathcal{X}_{u})), 
    \label{eq:general_unlearning}
\end{align}
where $\mathbb{D}(\cdot)$ denotes the distribution of a random variable.
For the problem of DGM where we consider the distribution of generated samples, and adapt Eq.~\eqref{eq:general_unlearning} into: 
\begin{align}
    \mathbb{D}(\mathcal{G}_{\mathcal{A}_\mathcal{U}(\mathcal{X}, \mathcal{X}_{u})}) = \mathbb{D}(\mathcal{G}_{\mathcal{A}(\mathcal{X} \backslash \mathcal{X}_{u})}),
    \label{eq:generative_unlearning}
\end{align}
where $\mathcal{G}_m$ is the generated samples from model $m$. By choosing different sub-dataset as $\mathcal{X}_{u}$, machine unlearning can fulfill various objectives. When $\mathcal{X}_{u}$ is set as the data owned and copyrighted by individuals, the DGM builder utilizes machine unlearning to safeguard the copyright of this data. \citet{zhang2023forget} pointed out four goals of unlearning for DGM: performance (successfully remove target data from the model), integrity (at best keep other data of the model), generality (can be applied to a wide range of data that covers all aspects of human perceptions) and flexibility (can be applied to various models of different tasks and domains). In the following, we discuss different unlearning methods for GANs and Diffusion Models in the image domain.

Fine-tuning with a modified objective is usually used to achieve Eq.~\eqref{eq:generative_unlearning} for DGMs efficiently. Compared with data filtering and re-training from scratch~\citep{nichol2022glide, ramesh2022hierarchical}, it is more efficient in time and energy. For example, a representative method \citep{kong2023data} modifies the min-max adversarial objective of GAN by mixing the undesirable data with generated data as negative (fake) samples of the discriminator.
Thus, the fine-tuned discriminator will not consider the undesirable data as true samples and the generator will be fine-tuned to avoid generating it.
Another unlearning method considers how to change the guidance of conditions in conditional DGMs, like text-to-image models. The copyrighted images can be transformed as concepts like styles, branches, person names, and so on. By re-directing the conditional representations of these concepts to conditions unrelated to the copyrights, the generated images can avoid infringement on the copyrighted images. \citet{kong2023data_conditional} represent the undesirable concepts as $\mathcal{C}_u$. They fine-tune the DGM to push the condition representation of concepts belonging to $\mathcal{C}_u$ towards a concept $\hat{c}$ which is not in the undesirable set. Meanwhile, to keep the other benign concepts unchanged, the fine-tuning objective also includes a term maintaining the representation of them as follows
\begin{align}
        \min _{H^{\prime}} L\left(H^{\prime} ; \lambda\right)= \mathbb{E}_{c \in \mathcal{C} \backslash \mathcal{C}_{u}}\left\|H^{\prime}(c)-H(c)\right\| + \lambda \cdot \mathbb{E}_{c \in \mathcal{C}_{u}}\left\|H^{\prime}(c)-H(\hat{c})\right\|,\label{eq:unlearn_pretrained_condition}
\end{align}
where $H^{\prime}(\cdot)$ is the fine-tuned condition representation function, while $H(\cdot)$ is the original condition function. The condition representation of undesired $\mathcal{C}_{u}$ is re-directed towards the benign concept $\hat{c}$ and can avoid to guide the model to produce the samples of concepts from $\mathcal{C}_{u}$. \citet{kong2023data_conditional} applied this general framework on different DGMs like class-conditional GAN~\citep{mirza2014conditional}, and GAN-based text-to-image models~\citep{zhu2019dm}. Besides, \citet{kongdiffwave} also extended this method to diffusion-based text-to-speech models. 


More unlearning methods by fine-tuning are tailored for Generative Diffusion Models, especially the text-to-image Stable Diffusion (SD)~\citep{rombach2022high} (refer to Section~\ref{sec:DGM}). Specifically,
\citet{kumari2023ablating} proposed to unlearn SD by matching the conditional distribution of undesirable concepts to anchor concepts in every diffusion step, which means the output of the denoising network $ {\epsilon}_{\theta}$ is flipped to anchor concepts that is unrelated to the undesirable ones. The anchor concepts can be random noise, $ {\epsilon}$, or the conditional distribution of benign concepts $\hat{c}$. With the anchor concept of random noise, the fine-tuning objective is 
\begin{align}
    \arg \min _{ {\epsilon}_\theta^{\prime}} \mathbb{E}_{ {\epsilon}, z, c \in \mathcal{C}_{u}, t}\left[\left\|  {\epsilon} - {\epsilon}_\theta^{\prime}\left( {z}_t, c, t\right)\right\|^2\right].
    \label{eq:unlearning_network_randomnoise}
\end{align}
This objective tries to induce the new diffusion network $ {\epsilon}_\theta^{\prime}$ to generate random noise if the input condition is from the undesirable concepts from $\mathcal{C}_{u}$.
If the model is unlearned by benign concepts, the fine-tuning objective is 
\begin{align}
    \arg \min _{ {\epsilon}_\theta^{\prime}} \mathbb{E}_{z, c \in \mathcal{C}_{u}, \hat{c}, t}\left[\left\|  {\epsilon}_\theta\left( {z}_t, \hat{c}, t\right) - {\epsilon}_\theta^{\prime}\left( {z}_t, c, t\right)\right\|^2\right].
    \label{eq:unlearning_network_desirable_set}
\end{align}
$ {\epsilon}_\theta\left( {z}_t, \hat{c}, t\right)$ is the output under benign conditions. By optimizing this objective, the output under undesirable concepts will be re-directed towards the benign concepts.
This method has a similar intuition to the second term in Eq.~\eqref{eq:unlearn_pretrained_condition}, but Eq.~\eqref{eq:unlearn_pretrained_condition} fine-tunes the component for conditional representation, while Eq.~\eqref{eq:unlearning_network_randomnoise} and Eq.~\eqref{eq:unlearning_network_desirable_set} fine-tune the whole diffusion network of SD. In addition, a similar regularization term to keep the other desirable concepts unchanged is also combined with Eq.~\eqref{eq:unlearning_network_randomnoise} and Eq.~\eqref{eq:unlearning_network_desirable_set}. 


\citet{zhang2023forget} proposed the method called Forget-Me-Not to resteer the cross-attention layers in SD by minimizing the values of attention maps, which disrupts the guidance of text condition in diffusion process.
In Stable Diffusion~\citep{rombach2022high, zhang2023forget}, the cross-attention of UNet transfers information from conditional text to hidden features through dot product and softmax. 
The attention maps are calculated based on hidden features and embeddings of condition texts. The attention maps will show the information of the undesirable concepts. Thus, distrupting the attention map will lead to a broken diffusion process. The fine-tuning objective of Forget-Me-Not minimizes the values in the attention map to disrupt the guidance of condition texts.
In this way, Forget-Me-Not can mislead and remove the undesirable concepts from the final generated data.

Different from the above two unlearning methods for diffusion models, \citet{gandikota2023erasing} targeted on the classifier-free diffusion generation~\citep{ho2022classifier}. 
The classifier-free diffusion guidance does not reply on a classifier for conditional generation, which is opposite to the classifier-based diffusion guidance. The classifier-based diffusion guidance modifies the denoising step to follow the feature captured by an auxiliary classifier $p_{\theta}(c|z_t)$:
\begin{align*}
    \tilde{{ {\epsilon}}}_\theta \left( {z}_t, {c}, t\right)= &  {\epsilon}_\theta\left( {z}_t, {c}, t\right) - w \nabla_{z_t} p_{\theta}(c|z_t),
\end{align*}
where $- \nabla_{z_t} p_{\theta}(c|z_t)$ provides the information of what features $z_t$ should have if $z_t$ wants to be classified as $c$. This guidance can shape $z_t$ into $c$ step by step, and $w$ is the parameter to control the strength of the classifier guidance~\citep{dhariwal2021diffusion}.
In contrast, the classifier-free diffusion guidance uses the difference between the outputs of conditional denoising diffusion model and unconditional denoising diffusion model as the guidance to boost the influence of conditions:
\begin{align*}
        \tilde{{ {\epsilon}}}_\theta \left( {z}_t, {c}, t\right)= &  {\epsilon}_\theta\left( {z}_t, {c}, t\right)+w(  {\epsilon}_\theta\left( {z}_t, {c}, t\right) -  {\epsilon}_\theta\left( {z}_t, t\right)),
\end{align*}
where $ {\epsilon}_\theta\left( {z}_t, {c}, t\right) -  {\epsilon}_\theta\left( {z}_t, t\right)$ is different between conditional and unconditional model, i.e. the influence of condition. It can also provide the information of the features that ${z}_t$ should possess to likely generate it into class $c$. The classifier-free diffusion model makes use of this information to guide the generation process.
\citet{gandikota2023erasing} proposed ESD which uses the opposite of the guidance as the unlearning objective to fine-tune SD:
\begin{equation*}
    \begin{aligned}
        \arg \min _{ {\epsilon}_\theta^{\prime}} \mathbb{E}_{\mathcal{E}(\mathbf{x}), c \in \mathcal{C}_{u}, t}  
        & \left[\left\| \Big ( {\epsilon}_{\theta^*}\left( {z}_t, c, t\right)-w({\epsilon}_{\theta^*}\left( {z}_t, c, t\right)  \right.\right. -  \left.\left.  {\epsilon}_{\theta^*}\left( {z}_t, t\right)) \Big ) - { {\epsilon}}_\theta^{\prime} \left( {z}_t, c, t\right)\right\|^2\right],
    \end{aligned}
\end{equation*}
The formulation uses several instances of diffusion models, $ {\epsilon}_{\theta^*}$ and $ {\epsilon}_{\theta}^\prime$. $ {\epsilon}_{\theta^*}$ is fixed for quantifying the opposite of guidance, while the $ {\epsilon}_{\theta}^\prime$ is the unlearned network to solve.
ESD has two variants, ESD-x and ESD-u. They have the same fine-tuning objective, but choose different subsets of parameters to fine-tune. ESD-x fine-tunes the cross-attention layers, while ESD-u fine-tunes non-cross-attention modules. \citet{gandikota2023erasing} found that the ESD-x can control the unlearned data according to the specific prompt, such as a named artistic style, and has little influence on other concepts. ESD-u can unlearn concepts from SD no matter whether a specific prompt is used, which is more suitable for global unlearning like NSFW.



\subsubsection{Dataset De-duplication}
\label{sec:data_duplication}

Memorization is another problem that threatens the copyright of source data. It refers to the problem that the generation model might produce images that are the same as the training data~\citep{van2021memorization}. It is found by many works that duplicated examples in the training data are more possible to be generated~\citep{webster2023reproducible, somepalli2023diffusion, carlini2023extracting}. \citet{somepalli2023diffusion} also argued that memorization is associated with the frequency at which data is replicated. \citet{carlini2023extracting} pointed out that the samples that are easy to be memorized usually duplicated for many times.

Therefore, dataset de-duplication is necessary to mitigate the memorization and prevent the model from generating the training data. It is attributed to an \textbf{active} method because de-duplication is conducted by the model builder for the purpose of providing legal services.
\citet{webster2023duplication} proposed a method to search the duplicate training images by CLIP~\citep{radford2021learning}. First, it trains the auto-encoders to compress the images and texts into latent space. It proposes Subset Nearest Neighbor CLIP (SNIP) to fine-tune the encoder for compression and keep the distance between neighbors at the same time.  After compression, it uses the inverted file system (IVF) to approximately search the duplicated images. The whole huge dataset is divided into small groups by $k$-means, and the duplication is searched within the closest centroids. Then the found data is de-duplicated to reduce the memorization.
OpenAI used a similar strategy to train DALL·E 2~\citep{openai_mitigate}. It first divides the training data into $K$ clusters, and then search the similar data within the union set of a small number of clusters. 

\citet{somepalli2023understanding} pointed out that the conditions of diffusion model, i.e. the text caption, will also influence memorization. Their experiments demonstrated that during the fine-tuning of Stable Diffusion, the memorization is more likely to happen if the captions (or text prompts) of different images are more diverse. But it does not mean the totally random captions can exacerbate memorization. Actually, the captions that are correlated with image content but have more diversity can lead to more severe memorization. However, diverse captions of the duplicated images can reduce the memorization. They also found that more training epochs have similar influence to duplication which increases the memorization. This means more epochs may mimic the duplication. Based on these findings, they concluded that for a single image and duplicated images, the diverse captions can help slow down the memorization. They proposed a few methods to process the captions during fine-tuning to avoid memorization, including multiple captions (which randomly samples from 20 captions generated by BLIP~\citep{li2022blip} for each image when training), random token replacement \& addition (which randomly replaces tokens/words in the caption with a random word, or add a random word to the caption at a random location) and so on.












\subsection{Model Copyright Protection}
\label{sec:model_protection}

This subsection summarizes the strategies for protecting the copyright of DGMs for image generation. This protection is important for two major reasons. First, a powerful DGM, which necessitates extensive computational  resources and well-annotated data for its creation, needs to be safeguarded from copyright infringements (such as being stolen by malicious users to offer unauthorized paid services). Second, the unregulated distribution of those models may lead to ethical concerns, including their potential misuse for generating misinformation, which necessitates techniques to identify the origin of an image.

Deep Generative Model Watermarking is a common solution for model copyright protection. It involves incorporating distinct information, known as a watermark, into the models before their deployment. The embedded watermark can be retrieved from a potentially infringing model or its generated data to confirm any suspected copyright violations. To achieve good performance in protecting the model while preserving the original generation performance, the watermarking technique should incorporate the following key properties: 
   (a) \textit{Fidelity}: the ability of the watermarking method to not significantly impact the general performance of the model (the diversity and visual quality of the generated images);
   (b) \textit{Integrity}: the accuracy with which the watermark can be extracted;
   (c) \textit{Capacity}: the length of message that can be effectively encoded to and extracted from the watermark;
   (d) \textit{Robustness}: the ability of the watermark to withstand alterations to the model or perturbation on the watermarked generated images;
   and (e) \textit{Efficiency}: the computational cost of the watermark embedding and extraction process.

The proposed DGM watermarking methods can be summarized into three categories according to the specific way they embed the watermark:
\begin{itemize}[noitemsep,topsep=0pt]
    \item \textbf{Parameter-based watermarking} encodes the watermark message into the model's parameters or structural configurations;
    \item \textbf{Image-based watermarking} embeds the watermark message into every image generated by the model;
    \item \textbf{Triggered-based watermarking} secretly incorporates a trigger to the protected model such that an image with copyright information will be generated once the trigger is activated.
\end{itemize}
 













\subsubsection{Parameter-based watermarking}

This type of watermarking techniques aims to subtly incorporate watermark information into the network's internal weights or structural configurations. It is known as a ``white-box" method because full access to these elements is required to implement and extract the watermark. Parameter-based watermarking has been widely applied for verifying the ownership of classification models \citep{uchida2017embedding,darvish2019deepsigns,zhao2021structural}  and has also been extended to protect DGMs in recent years. 

\citet{ong2021protecting} proposed to embed the signature of a GAN model to the scaling factors, $\gamma$, of the normalization layer of the generators. The loss for watermark embedding is:
\begin{align}
\mathcal{L}_s(\gamma, {B})=\sum_{i=1}^C \max \left(\gamma_0-\gamma_i b_i, 0\right),
\label{eq:parameter_watermark}
\end{align}
where $B=\left\{b_1, \cdots, b_C \mid b_i \in\{-1,1\}\right\}$ refers to the predefined binary watermark message and $C$ denotes the number of channels. This objective stipulates that the scaling factor of the $i$-th channel, denoted as $\gamma_i$, adopts either a positive or negative polarity (+/-) as determined by $b_i$. The term $\gamma_0$ is a constant employed to regulate the minimum value of $\gamma$. The sign loss is added to the original training loss of the GAN model (e.g., a  DCGAN \citep{radford2015unsupervised}, SRGAN \citep{ledig2017photo} or CycleGAN \citep{zhu2017unpaired}) to form the overall training objective of the generators. With this embedding strategy, the capacity of the watermark is determined by the total number of channels in the normalization layers. In the verification stage, the embedded message can be easily extracted by looking at the signs of the specific scaling factors of the model.


\subsubsection{Image-based watermarking}
This group of watermarking approaches embeds watermark messages into all the images generated by the model to be protected.  To achieve this goal, various studies suggested using an additional deep-learning-based network for embedding watermarks after the image generation process. In detail, once the DGMs create the images, these extra watermark embedding networks apply watermarks to the images before they are released to public.
For example, \citet{zhang2022generative} proposed two frameworks for watermark embedding based on the Human Visual System (HVS) to reduce the impact of the watermark on the visual quality of the generated images. Leveraging the fact that humans are more sensitive to changes in green color and brightness, they recommended embedding watermarks into the Red (R) and Blue (B) channels in the RGB color system or the U/V channels in the Discrete Cosine Transform (DCT) domain.
In either framework, each channel carries a unique watermark, allowing for the extraction of two distinct watermarks from a single image. 

\begin{wrapfigure}{r}{0.5\textwidth} 
    \centering
    \vspace{-0.18in}
    \includegraphics[width=\linewidth]{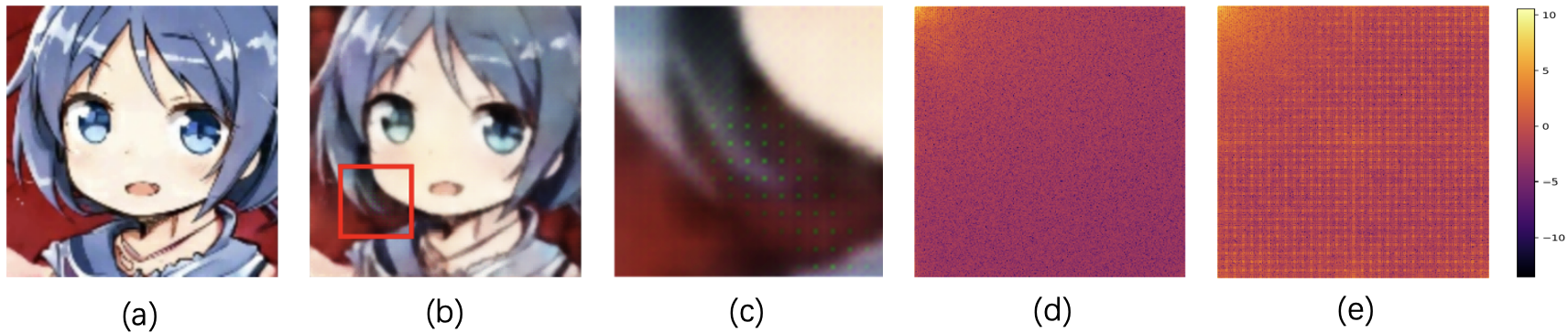}
    \vspace{-0.3in}
    \caption{\small  High-Frequency Artifacts.  (a) ground-truth image, (b) watermarked image by \citet{wu2020watermarking}, (c) partial zoom-in of (b), (d)  DCT heat maps of (a), (e) DCT heat maps of (b). 
    (\citet{zhang2023generative})}
    \label{fig:artifacts}
    \vspace{-0.1in}
\end{wrapfigure}
As a following-up work, \citet{zhang2023generative} noted that images marked using prior watermark embedding networks display significant high-frequency artifacts in the frequency domain. As shown in Figure~\ref{fig:artifacts},  slight spatial artifacts are detectable in the generated image marked by the method described in \citet{wu2020watermarking}. And obvious grid-like high-frequency
artifacts can be found in the DCT heat map of the marked image. These artifacts, mainly due to the up-sampling and down-sampling convolution operations of the GAN-like embedding network, could compromise the watermark's imperceptibility.  To tackle the problem, they designed a new watermark embedding network which is capable of suppressing high-frequency artifacts through anti-aliasing. The anti-aliasing is primarily implemented by introducing a low-pass filter before the down-sampling process and appending a convolutional layer after the nearest-neighbor up-sampling process, adopting strategies demonstrated to be effective in previous research.
In addition to using a separate watermark embedder, some other strategies consider directly modifying the host generative network so that the host network itself can embed the watermark while achieving the original generation task. At the same time, an external watermark decoder should be trained to correctly extract the watermark information from the generated images. To achieve this goal in GANs, \citet{fei2022supervised} proposed to update the training objective of the generator by:
\begin{align*}
    \begin{split}
        L_G^w(D, G) =  \mathbb{E}_{{z} \sim p_{z}}[\log (1-D(G({z})))] + 
         \gamma \mathbb{E}_{{z} \sim p_{z}}[BCE(D_w(G({z})), w_{gt})],
    \end{split}    
\end{align*}
where $D$ and $G$ refer to the discriminator and generator of GAN respectively, $D_w$ refers to the watermark decoder, $p_z$ means the prior distribution of the latent space, $w_{gt}$ is the ground truth watermark message, $\gamma$ is a regularization parameter and $BCE$ represents the binary cross entropy. The first term in the equation is the standard generator loss, while the second term is for ensuring the correctness of the decoded binary watermark message. 
It is noted that the watermark decoder $D_w$ they applied here is retrieved from a well-developed standard image-watermarking framework and its parameters are fixed in the training process of $G$. With the above training objective, the model can be either trained from scratch or fine-tuned from a non-watermarked pre-trained GAN model.


\citet{yu2020responsible} also proposed a fingerprinting technique to trace the outputs of GANs back to their source, which can help identify misuse. The core difference between it and the work of \citet{fei2022supervised} is that the trained generator $G$ takes both the latent code ${{z}}$ and the embedding of the fingerprint $\boldsymbol{c}$ (a sequence of bits) as its input, which increases the efficiency and scalability of the fingerprinting mechanism. Once the model is trained, multiple instances of fingerprinted generators which focus on embedding different fingerprint messages can be directly obtained. In their algorithm, the watermark encoder, $E$, which maps fingerprint $\boldsymbol{c}$ to its embedding, the decoder, $F$, which obtains the latent code ${z}$ and fingerprint $\boldsymbol{c}$ from a watermarked image, the discriminator $D$ and the generator $G$ of GAN are optimized together with the following loss:
\begin{align*}
\min _{E, F, G} \max _D \lambda_1 \mathcal{L}_{a d v}+\lambda_2 \mathcal{L}_z+\lambda_3 \mathcal{L}_c+\lambda_4 \mathcal{L}_{\text {const}}.   
\end{align*}

In this objective, the first term $\mathcal{L}_{a d v}$ refers to the original training loss of a GAN model and the second term is the 
regularization term firstly proposed by \citet{srivastava2017veegan} for mitigating the mode collapse issue of GANs. The term $\mathcal{L}_{c}$
is the cross entropy loss of the fingerprint decoding to ensure the correctness of the reconstruction of the fingerprint.  The $\sigma\left(\cdot\right)$ here refers to the sigmod function.
The formulation of the last term  $\mathcal{L}_{\text {const}}$ is
designed to ensure the perceptual similarity between images generated with the same latent code but different fingerprints, which guarantees that the latent code exclusively controls the content of the generated images, irrespective of the fingerprint variations. 

Similar idea of weights modulation has also been applied for the protection of LDMs.
\citet{kim2023wouaf} proposed to incorporate the watermark message to generated images by modulating the parameters of each layer of the decoder $\mathcal{D}$ of LDM. Specifically, 
after the watermark message $\phi$ is obtained, a mapping network $\mathcal{M}$ is applied to derive the feature representation of the message. Then an affine transformation layer $\mathcal{A}_l(\cdot)$ is developed for each layer $l$ of the decoder $\mathcal{D}$. The affine transformation layer is applied to match the dimension of the message representation to the dimension of the model's weights. Then the weight modulation is conducted following the formulation:
\begin{equation}
   W^{\phi}_{i,j,k}=u_j*W_{i,j,k}, 
   \label{eqn:weight_modu}
\end{equation}
where $W$ and $W^\phi$ refers to the pre-trained and fingerprinted
parameters of the decoder $\mathcal{D}$, $i$, $j$, $k$
denote the dimensions of the input, output, and kernel of each layer, and $u_j=\mathcal{A}_l(\mathcal{M}(\phi))$ denotes the scale of the modulation to the $j$th output
channel. When the watermark message is encoded, a watermark decoder $\mathcal{F}$ is also required to retrieve the watermark from the generated images. 
The watermark decoder $\mathcal{F}$ as well as the mapping network $\mathcal{W}$ are jointly trained with the following loss:
$$\min_{\mathcal{A},\mathcal{M},\mathcal{D}, \mathcal{F}} \lambda_1 L_{\phi} + \lambda_2 L_{\text{quality}},
$$
where $L_{\phi}$ denotes the
binary cross entropy of watermark decoding:
\begin{align*}
L_{\phi} &= \mathbb{E}_{z=\varepsilon(x), \phi \sim \Phi}  \sum_{i=1}^{d_{\phi}} \big( \bigg[\phi_i \log \sigma\mathcal{F}(\mathcal{D}(\phi, z))_i + (1 - \phi_i) \log(1 - \sigma(\mathcal{F}(\mathcal{D}(\phi, z)))_i \big) \bigg],
\end{align*}
in which $z=\varepsilon(x)$ refers to the latent feature of an image $x$,
$\mathcal{D}(\phi, z)$ denotes the output of the decoder whose weights being modulated by Eq.(\ref{eqn:weight_modu}), and $\sigma$ refers to the sigmoid activation function. $\Phi$ refers to a Bernoulli distribution. $L_{\text{quality}}$ is the regularization term inhibiting the influence of the watermark to the quality to the generated image.

Another work~\cite{xiong2023flexible}, which also aims to protect the copyright of LDMs, has a similar pipeline to \citet{kim2023wouaf}. The major difference is that the feature representation of the watermark message is not used to modulate the weight of the decoder $\mathcal{D}$, but is combined with the intermediate outputs of the fine-tuned decoder $\mathcal{D}$ such that the image generated by $\mathcal{D}$ is watermarked. Under the pipeline, a watermark decoder is also required to extract the watermark from the generated image.

Additionally, the work of \citet{fernandez2023stable}, which also focus on the protection of Latent Diffusion Models (LDMs), suggested to directly apply a classical neural network-based watermarking method, HiddeN~\citep{zhu2018hidden}, to jointly optimize the parameters of a watermark encoder and extractor. Then the decoder of the LDM is fine-tuned such that all images it generates contain a given watermark that can be extracted by the pre-trained watermark extractor. 

Unlike the previously mentioned methods that require modifying the generative networks to include watermarks in their generated images,  \citet{yu2021artificial} achieved this by solely watermarking the training images of the generative models.  To ensure that the watermark can be successfully transferred from the training images to the generated images of GANs, they developed a deep-learning-based structure for watermark (or ``fingerprints'' in their paper) embedding and extraction. Similar to the other watermark embedding-extraction framework,
the training objective of the watermark embedder and decoder also consists of a $BCE$ loss which guides the decoder to decode the fingerprint correctly and an $MSE$ loss which penalizes any deviation of the watermarked image from the original image. Although the framework proposed by \citet{yu2021artificial} was originally designed for DeepFake detection and misinformation prevention, its functionality can be extended to verifying the copyright of GANs, as the watermark on the generated images can indicate their source. Additionally, \citet{zhao2023recipe} experimentally verified that this technique can be extended to safeguard the copyright of unconditional and class-conditional diffusion models.

\begin{wrapfigure}{r}{0.4\textwidth} 
    \centering
    \vspace{-0.18in}
    \includegraphics[width=\linewidth]{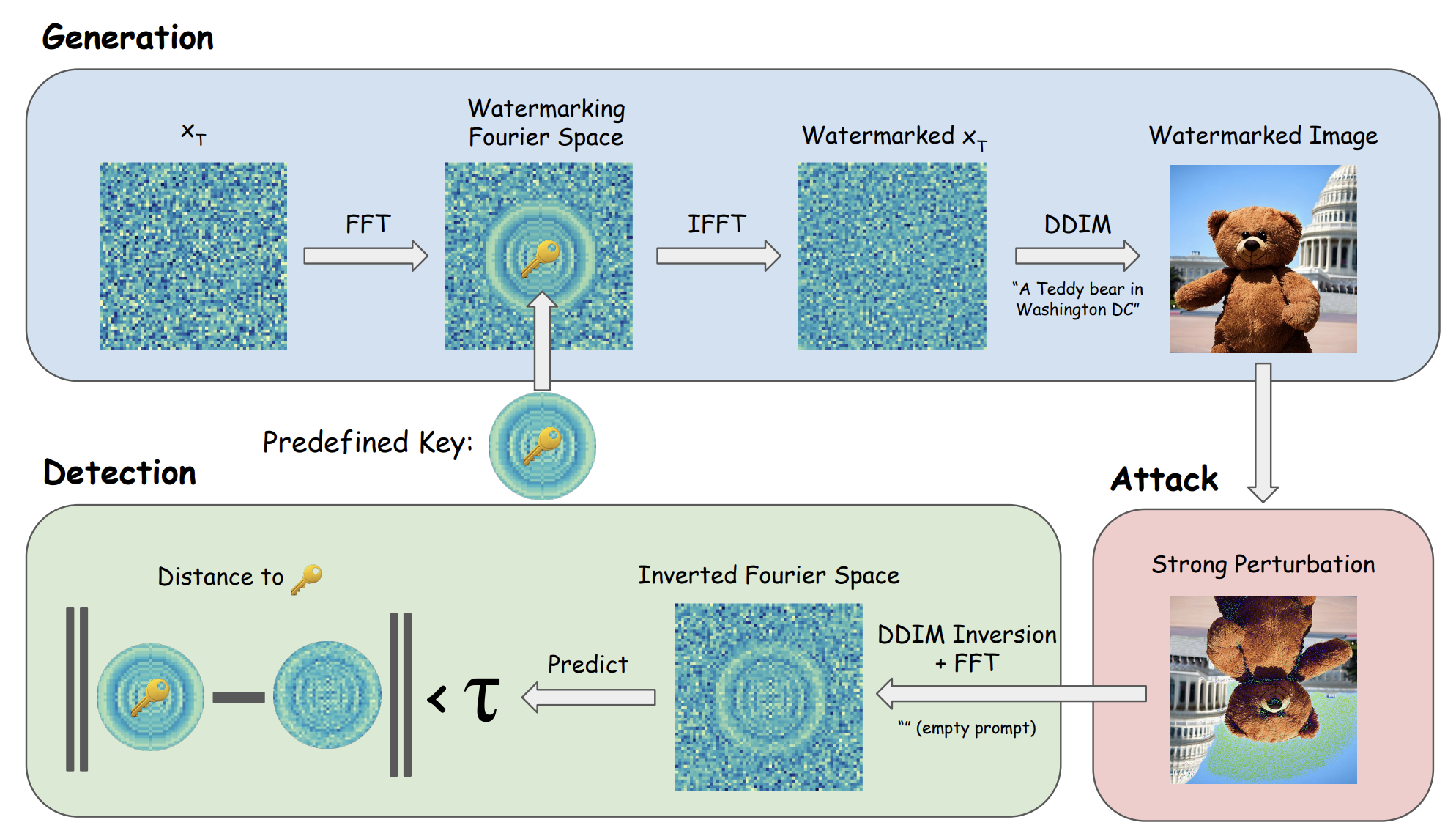}
    \vspace{-0.3in}
    \caption{\small  Pipeline for Tree-Ring Watermarking. (\citet{wen2023treering})}
    \label{fig:tree_ring}
    \vspace{-0.1in}
\end{wrapfigure}

While the previous methods consider incorporating the watermarking embedding procedure in the training process of the generative models,  \citet{wen2023treering} proposed Tree-Ring Watermarks, a watermarking framework specified for diffusion models which conducts the watermark encoding in the sampling process. 
As shown in Figure~\ref{fig:tree_ring}, the watermark is embedded into the initial noise vector used for sampling. In order to ensure that the watermark can achieve a better robustness against multiple image modification such as cropping, dilation, flipping, and rotation, they suggest to encode the watermark patterns to the Fourier space of the image.
After the image is generated, watermark detection is done by inverting the diffusion process to reconstruct the noise vector. This is done using the DDIM \citep{song2020denoising} inversion process. In detail, the initial noise vector $\boldsymbol{x}_T \in \mathbb{R}^L$ is described in Fourier space as:
\begin{align*}
\mathcal{F}\left(\boldsymbol{x}_T\right)_i \sim \begin{cases}k_i^* & \text { if } i \in M \\ \mathcal{N}(0,1) & \text { otherwise. }\end{cases}
\end{align*}
where $M$ refers to a binary mask, $k^* \in \mathbb{C}^{|M|}$ is the key chosen for the watermark. The key is designed to be ring-shaped in the Fourier space to ensure that the watermark is invariant to certain common image transformations. In the detection stage, the watermark is detected if the $L_1$ difference between the inverted noise vector and the pre-defined key in the Fourier domain of the watermarked area $M$ is below a tuned threshold $\tau$.

Another work which considers adding the watermark in the images' sampling process is~\citet{nie2023attributing}. This approach, which target to protect the copyright of LDM or StyleGAN2~\cite{karras2020analyzing}, does not require a network to conduct the watermark embedding and decoding but suggest to directly modify the latent features $\psi(z) \in \mathbb{R}^{d_w}$ (either generated by the multi-layer perception network of the StyleGAN2 or sampled by the diffusion process of LDM) by matrix operation. In detail, let $U\in \mathbb{R}^{d_w\times (d_w - d_{\phi})}$ denotes an orthonormal subspace of the space of the latent feature $z$ and $V \in \mathbb{R}^{d_w\times d_{\phi}}$ denotes its complementary subspace, the watermarked latent feature is 
\begin{align*}
    w_{\phi}(\alpha) = U\alpha + \sigma V\phi, \text{\quad where }\alpha = U^{\dagger} \text{proj}_U \psi(z) \in \mathbb{R}^{d_w - d_{\phi}}.
\end{align*}
The notation $\phi \in \mathbb{R}^{d_\phi}$ refers to the watermark message, $U^{\dagger}$ represents the pseudo-inverse of $U$, $\text{proj}_U \psi(z)$ denotes the projection of $\psi(z)$ to $span(U)$, and $\sigma$ denotes a hyperparameter to control the strength of the watermark.
In the detection stage, in order to retrieve the watermark from the image, an optimization problem is solved:
\begin{align*}
\min_{\hat{\alpha},\hat{\phi}} \ell \left( g(w_{\hat{\phi}}(\hat{\alpha})), g(w_{\phi}(\alpha)) \right) \quad \text{s.t.}~\hat{\alpha}_i \in [\alpha_{l,i}, \alpha_{u,i}],~ \forall i = 1, \ldots, d_w - d_{\phi},
\end{align*}
where $l$ refers to the LPIPS metric~\cite{zhang2018unreasonable} which evaluates the visual similarity of two images. The lower and upper bound of $\hat{\alpha}_i$, $\alpha_{l,i}$ and  $\alpha_{u,i}$, are chosen based on empirical observation.  


    



\subsubsection{Trigger-based watermarking} 
\label{sec:image_trigger_based}

The trigger-based watermarking follows the basic technical scheme of backdoor attack~\cite{wang2019neural, saha2020hidden, chen2017targeted}, which embeds a trigger into a neural network to cause a failed classification by activating the trigger. In the protection of DGM copyright, once the trigger is integrated into the protected model, the activation of this trigger in the input will cause the model to generate a watermarked output. By examining the presence or absence of the watermark in this output, the model owner can determine whether a suspect model has been illicitly derived from the protected, watermarked model. Consequently, if the watermark is detected in the example of triggered generation, it substantiates the claim of copyright infringement regarding the model.

\citet{ong2021protecting} also proposed a trigger-based watermark framework to protect the copyright of GANs. It first adds the trigger onto $z$ which is the input of the generator $G(\cdot)$ to get the triggered input $z_{\text{trigger}}$. The protected GAN is trained towards the goal that once the trigger is input to $G(\cdot)$, the output will be generated with a watermark. The model owner can input a trigger into the suspect model and verify whether the model is stolen from the copyrighted model by checking the existence of watermark in the output. For different types of GAN, the triggered input is obtained by different manually defined rules. For example, in DCGAN, the trigger is encoded with a binary representation ${b}$:
\begin{align*}
 z_{\text{trigger}} = {z}\odot {b}+c\left( 1-{b} \right) ,{b}\in \left\{ 0,1 \right\} ^{d_z},  
\end{align*}
where $\odot$ is the elementwise production, and ${d_z}$ is the dimension of $z$. The triggered input $z_{\text{trigger}}$ means each dimension of $z$ is encoded with one dimension in $b$.
The triggered input is expected to cause $G(\cdot)$ to generate a watermarked target $x_{\text{target}}$. During the training of GAN, the trigger pair~$\left(z_{\text{target}}, x_{\text{target}}\right)$ is included in a regularization term for this. The term is formulated as
\begin{align*}
\mathcal{L}_w\left(z_{\text{trigger}}, x_{\text{target}}\right)=1-\operatorname{SSIM}\left(G\left(z_{\text{trigger}}\right), x_{\text{target}}\right),
\end{align*}
where SSIM refers to the structural similarity. The regularization term $\mathcal{L}_w$ is combined with the training loss of $G(\cdot)$ by a coefficient $\lambda$:
\begin{align*}
    \mathcal{L}_{\mathrm{G}_w} = \mathcal{L}_{\mathrm{G}}+\lambda \mathcal{L}_w.
\end{align*}

However, the trigger-based method is not implemented in isolation. Instead, it is integrated with the parameter-based watermarking approach, as detailed in Equation~\eqref{eq:parameter_watermark}, following the practice outlined in~\cite{ong2021protecting}.

Besides the protection of GANs, \citet{liu2023watermarking} proposed to watermark Stable Diffusion by injecting triggers into the prompt. According to the method, when specific triggers are present in the prompt, Stable Diffusion will produce an image with a watermark. If the resulting image closely resembles the watermarked prototype, it confirms that the model is safeguarded by the designated watermark. \citet{liu2023watermarking} developed two different approaches for integrating triggers, NaiveWM and FixedWM. NaiveWM embeds a specific word at a random location within the prompt, while FixedWM places a specific word at a predetermined position in the prompt. The chosen trigger word should be nonsensical to avoid accidental activation. For embedding the trigger-based watermark into Stable Diffusion, a triggered dataset is employed to fine-tune the model. Additionally, a clean dataset is also used in the fine-tuning process to preserve the quality of generation when the trigger is not activated. \citet{zhao2023recipe} proposed a similar idea, but did not consider the stealthiness of the trigger. Their method directly uses ``[V]'' as the whole triggered prompt.

Different from \citet{liu2023watermarking} and \citet{zhao2023recipe} that embedded the watermark into the guidance of prompt, \citet{peng2023protecting} watermarked by injecting triggers into the diffusion process. In the reverse process of generative diffusion models, when the trigger is added onto the current step, the subsequent will be generated towards the watermarked image. By repeating injecting the trigger into the reverse process, the final generated image will be a watermarked image that can be used to verify the copyright of the model. For embedding the watermark, the model is trained with both triggered dataset and clean dataset. During training, when the data comes from triggered dataset, the triggered noise is input into the denoising network of diffusion model and the network is optimized to denoise towards the watermarked image. During the both forward and reverse process, the trigger is added into step $t$ by
\begin{align*}
    {x}_t^{\text{trigger}} = \gamma_1 {x}_t+\left(1-\gamma_1\right) {b},
\end{align*}
where $b$ is the trigger, and ${x}_t$ is the state of step $t$. For reverse process, ${x}_t$ is the image denoised from the last step, while for forward process, it is the diffused image accumulated by the Gaussian noise in the forward steps. This protection method can be used for both training from scratch and fine-tuning. Although this method can also protect the copyright by verifying whether the final output is watermarked or not, the extraction of watermark requires the modification on each diffusion step which is more strict than the previous methods that only need to trigger in the prompt.

\section{Copyright in Text Generation}\label{sec:llm}

In this section, we first define the problem of copyright protection in text and unique properties in text domain. Then we introduce data copyright and model copyright protection by providing the taxonomy of existing methods.

\subsection{Background: DGMs for text generation}\label{sec:llm_def}



In the realm of text generation, our focus is primarily on Large Language Models (LLMs), due to their outstanding capabilities and the potential risks they entail. As key components of DGMs, LLMs have notably propelled advancements in the field of text generation. They exhibit remarkable emergent abilities, significantly boosting performance in a range of NLP tasks. However, alongside their exceptional efficacy, LLMs also present substantial concerns, particularly in the realms of copyright. The considerable commercial value and the high costs associated with their training further underscores the importance of addressing these issues. In Section~\ref{sec:llm}, which focuses on the text domain, we explore LLMs in detail, acknowledging their unparalleled efficiency and effectiveness in text generation, while also considering the associated concerns in copyright protection.

\textbf{Training and Inference of LLMs.} Generally, Large Language Models (LLMs) are constructed to comprehend human language and produce coherent, contextually relevant text. To achieve this objective, most existing LLMs can be categorized into ``\textit{pre-trained LLMs}'' and ``\textit{fine-tuned LLMs}''. For the pre-trained LLMs, they are trained on a massive amount of internet text data, aiming to grant them the ability to predict the probable subsequent token based on its preceding tokens. Through this process, the models can obtain an understanding of the patterns and structures inherent in human language. Many early LLMs, such as GPT-2~\cite{radford2019language}, GPT-3~\cite{brown2020language}, OPT~\cite{zhang2022opt}, GPT-Neo~\cite{gpt-neo}, and others, have been developed in this procedure, to be equipped with the ability to complete sentences by generating the possible subsequent tokens.

Through various fine-tuning technologies, LLMs have been made increasingly flexible and adept at various downstream tasks. Specifically, the Instruction Tuning~\cite{zhang2023instruction} strategy is employed to enable LLMs to handle diverse language tasks based on users' requests. This versatility allows LLMs to be applied as general assistants for tasks such as question answering~\cite{lu2022learn}. Furthermore, Reinforcement Learning with Human Feedback (RLHF) is utilized to help LLMs better align with human values, enhance the reliability of generated outputs, and improve ethical decision-making~\cite{ouyang2022training}. Based on these advancements, advanced LLMs like ChatGPT~\footnote{\href{https://chat.openai.com/}{https://chat.openai.com/}}, Claude\footnote{\href{https://www.anthropic.com/index/claude-2}{https://www.anthropic.com/index/claude-2}}, Bard~\cite{manyika2023overview}, and LLaMA 2~\cite{touvron2023llama}, have been developed.

\textbf{Notations.} Next, we introduce key notations we will use in this paper. We denote an arbitrary text $x$ which is composed of a sequence of tokens $(w_1, w_2, ..., w_k)$ with length $k$. Many popular LLMs follow an ``auto-regressive'' manner: given a sequence of prior tokens $(w_1, w_2, ..., w_{i-1})$, a language model $F$ calculates the probability of the next token $\hat{w_{i}}$ conditioning on the preceding tokens. We denote the probability that model predict the $i$-th word as $v_j$ by $p_F(\hat{w_i} = v_j | w_1, ..., w_{i-1})$, where $v_j \in \mathcal{V}$ and $\mathcal{V}$ is the vocabulary. Based on this design, the  \textbf{\textit{likelihood}} of each token in the original sentence $x$ given by the model $F$ can be defined as: 
\begin{equation}\label{eq:ps}
p_F(\hat{w_i} =\ w_i | w_1, ..., w_{i-1}) 
\end{equation}
which is the probability of the model to give an output token to be the same as $w_i$ in the text. Similarly, we can also define the likelihood of the whole sentence $x$ given by the model $F$, which is the overall likelihood of all tokens in the sentence: 
\begin{equation}\label{eq:pss}
p_F(x) = \prod^k_{i=2} p_F(\hat{w_i} =\ w_i | w_1, ..., w_{i-1}) 
\end{equation} 
For most existing LLMs, during the (pre-)training process, the model is trained to maximize this likelihood, so that the model can learn to generate the texts following the distributions of training texts. During the LLM's inference process, the model will generate the texts, by making sampling from the tokens with high likelihood.

\subsection{Copyright Issues in Text Generation}

\textbf{Data copyright protection.}
The data copyright problem regarding the use of LLMs has attracted  extensive focus and debates. 
The definition of copyright violation varies across countries and laws. Generally, it means \textit{the use of works protected by copyright without permission for a usage where such permission is required, thereby infringing certain exclusive rights granted to the copyright holder, such as the right to reproduce, distribute, display or perform the protected work, or to make derivative works}\footnote{\href{https://en.wikipedia.org/wiki/Copyright_infringement}{https://en.wikipedia.org/wiki/Copyright\_infringement}}.  Similarly, 
\citet{lee2023language} also note that \textit{``plagiarism occurs when any content including text, source code, or audio-visual content is reused without permission or citation from an author of the original work."} However, as evident from the following laboratory and court cases, current LLMs demonstrate various plagiarism behaviors, which can be divided into three types: verbatim plagiarism, paraphrase plagiarism, and idea plagiarism:
\begin{itemize}[noitemsep,topsep=0pt]
    \item \textit{Verbatim Plagiarism}, which refers to directly copying the origin data completely or partially.
    \item  \textit{Paraphrase Plagiarism}, which refers to composing new works including consent word replacement, statement rearrangement, or even translate sentences back and forth, from copyrighted works. 
    \item \textit{Idea Plagiarism}, which refers to copying the core idea of the copyrighted material.
\end{itemize}

\textit{Cases in Laboratory.}
There are existing lines of work showing that Large Language Models (LLMs) tend to memorize and emit parts of its training data (which may include copyrighted materials), referred to as the ``memorization effect of LLMs''. Although such effect is widely believed to be essential for language model's performance and generalization, it also raises serious risks on data copyright protection. \citet{carlini2021extracting} first found that LLMs can memorize and leak specific training examples, by devising a data extraction attack, which can effectively examine the training data verbatim from GPT-2. Then, \citet{carlini2022quantifying} further quantified such phenomena in detail and found that the memorization effect grows with the  model scales, number of  replicates, as well as the prompt length. Similarly, \citet{tirumala2022memorization} suggested that lager LMs generally memorize faster and less likely to forget information. Besides, they found that nouns and numbers are more likely to be memorized. 

\citet{zhang2021counterfactual} defined "counterfactual memory" which quantifies the performance difference between models trained on the specific data and models not trained on that. \citet{biderman2023emergent} investigated correlations of memory phenomena between large and small models and between partially-trained and fully-trained models. They suggested that we can use partially trained-model to efficiently predict whether the data is memorized by a fully-trained model. Furthermore, \citet{zeng2022exploring} explored the memorization behavior of fine-tuned LLMs and identified some feature dense tasks such as dialog and summarization present high memorization effects, and demonstrated the correlation between attention scores and task-specific memorization. 

\textit{Cases in Court.}  Recently, The New York Times initiated a lawsuit against OpenAI and Microsoft, alleging copyright infringement\footnote{\href{https://www.nytimes.com/2023/12/27/business/media/new-york-times-open-ai-microsoft-lawsuit.html}{https://www.nytimes.com/2023/12/27/business/media/new-york-times-open-ai-microsoft-lawsuit.html}}. The lawsuit asserts that OpenAI and Microsoft used millions of articles from The Times to train their automated chatbots. These chatbots, the suit contends, now rival The Times in providing reliable information. The complaint seeks restitution for what The Times describes as ``billions of dollars in statutory and actual damages'' arising from the ``unauthorized replication and exploitation of The Times’s distinct and valuable works.'' A specific allegation made in the lawsuit is that ChatGPT, when queried about current events, occasionally generates responses containing ``verbatim excerpts'' from articles published by The New York Times. These articles are typically behind a paywall, accessible only to subscribers. Moreover, the lawsuit highlights instances where the Bing search engine, incorporated with ChatGPT, reportedly displayed content sourced from a New York Times-owned website. This usage was done without providing direct links to the articles or including the referral links that The Times employs for revenue generation.

\textbf{Model copyright protection.} 
For LLMs, the infringement on model copyright is in line with the model copyright protection in the image domain. The model builder invests a significant amount of funds and labor into the construction of LLMs, which naturally grants them intellectual property rights over the trained model.

\subsection{Data Copyright Protection}\label{sec:llma}

According to the previous discussed cases, a majority of research works study how to improve the training scheme of LLMs to avoid replicating its training data samples. These methods are generally proposed from the perspective of model builders to prevent copyright infringement and provide legal services. In text domain, they can be categorized according to the major reasons which cause the memorization / plagiarism behaviors of LLMs:
\begin{itemize}[noitemsep,topsep=0pt]
    \item \textbf{Data-deduplication}, which is the strategy of removing the repeated samples from the training set of LLMs. As evident from existing studies such as ~\citep{lee2021deduplicating, kandpal2022deduplicating}, the repeated contents from the training set are easier to be memorized and reproduced by LLMs. 
    \item \textbf{Improved training \& generation algorithms}: the model builder can also modify the training objective and text generation procedure to avoid potential reproduction behavior of LLMs.
    \item \textbf{Alignment strategies}: the model builder devises new alignment strategies to reduce memorization in LLMs.
    \item \textbf{Machine unlearning}, which services to delete the copyrighted materials from LLMs, once the owners of the copyrighted materials identify the infringement.
\end{itemize}
Similar to image domain, machine unlearning can be seen as the passive method requested by the data owner, while the others are active approaches directly conducted by the model builder during the model construction stage.

\subsubsection{Data De-duplication}\label{sec:llm1}

In the training of Large Language Models (LLMs), a key issue is the memorization of training data $\mathcal{X}$, particularly when it involves copyrighted data $\mathcal{C} \subset \mathcal{X}$. To improve copyright security, one major method is to {de-duplicate to reduce memorization}. It refers to the process that the model builder removes the duplicated data samples from the training set of LLMs. \citet{lee2021deduplicating} found that existing NLP datasets contain many duplicated and near-duplicated substrings, leading to over 1\% of language model outputs being exact copies from training data. To address this, they developed ExactSubstr and NearDup for dataset de-duplication. ExactSubstr removes long shared substrings between two samples by concatenating the entire dataset into a single sequence and using a suffix array for easy deletion of adjacent repetitive sequences. NearDup utilizes MinHash \cite{broder1997resemblance} to identify potential document matches, followed by rigorous similarity checks or direct deletion of duplicates. This process, which includes sorting $n$-grams of each document and applying the Jaccard Index for similarity, significantly reduces model memorization, decreases training time, and improves evaluation accuracy by reducing train-test overlap.  \citet{kandpal2022deduplicating} found that language models' likelihood of regenerating training sequences superlinearly correlates with their frequency in the dataset. They also noted significant improvements in privacy protection following de-duplication in these models.


\subsubsection{Improved Training \& Generation algorithms}\label{sec:llm2}
\citet{chu2023protect} define copyright protection for LLMs as ensuring the model's output for any input does not significantly resemble any copyrighted content in $\mathcal{C}$, as measured by a metric $L$. The training goal for a dataset $\mathcal{X}$ with copyrighted and non-copyrighted data is to fine-tune the model $f$ to achieve:
\begin{equation}\label{eq:similarity}
    L(f(x), \mathcal{C}) \geq \tau + L(f(x), \mathcal{X}_{\neg\mathcal{C}}),
\end{equation}
where $\mathcal{X}_{\neg\mathcal{C}} = \mathcal{X} \cap {\neg\mathcal{C}}$, $L$ measures the possibility to generate $\mathcal{C}$ and higher $L$ means less possibility of generating $\mathcal{C}$. Thus, Eq.~\eqref{eq:similarity} means the output of $f$ should maintain less possibility to come from copyrighted data or be similar to data.
To attain this goal, \citet{chu2023protect} introduce a method known as copyright regression. In a simplified regression case, $(A, b)$ denote training data, where $A$ is the input and $b$ is the target output. The model is trained to get $f(A)=b$. $(A_1, b_1)$ denote copyrighted data. An additional term \(L(f(A_1), b_1)^{-1}\) is added to the training objective to discourage the model from generating outputs matching the copyrighted data, with a scalar coefficient \(\gamma > 0\). Consequently, the training objective is modified to \(L(f(A), b) + \gamma L(f(A_1), b_1)^{-1}\), integrating an inverse term that strikes a crucial balance between performance and copyright protection, and helps mitigate the model's tendency to generate copyrighted outputs. However, the method in~\citep{chu2023protect} requires the model builder to know which training samples are copyrighted, which are not executable for most existing LLMs.

In the text generation stage, there are also ways to protect copyright, even if the model has memorized a part of copyrighted data. These methods manipulate the generation process of LLMs to avoid the occurrence of copyrighted information. 
\citet{ippolito2022preventing} developed a memorization-free (MEMFREE) decoding strategy using bloom filters to effectively eliminate all direct memorization of text. Unlike traditional retroactive censoring, which repetitively runs the language model with varied seeds but identical prompts until non-resembling training set outputs are generated, MEMFREE efficiently targets memorization at the $n$-gram level during the decoding phase, rather than examining the entire sentence. Specifically, it assesses whether any $n$-gram in the generated sequence matches those in the training set, suppressing any such occurrences and prompting re-sampling from the model. This approach, however, fails in the case that the generated text is similar but not exactly the same as the memorized content.  The authors highlight that strict definitions of verbatim memorization are insufficient, as they do not address subtler forms of memorization. They note that even models adept at avoiding exact memorization are vulnerable to approximated memorization, which can be subtly bypassed through creatively altered "style-transfer" prompts.

\subsubsection{LLM Alignment Protection}\label{sec:llm5}

Refer to the discussion in Section~\ref{sec:llm_def}, LLM alignment strategies, such as RLHF, have been widely applied for constructing better LLMs aligned with human
values, such as enhancing the reliability of generated outputs, and improving ethical decision-making. Therefore, the similar pipeline of model alignment can also be investigated from the perspective of copyright protection. The study by \citet{kassem2023preserving} introduces a model alignment strategy designed to prevent memorization in LLMs. Namely, they proposed
a reward-based DeMemorization (DeMem) framework to reduce memorization in language models (LMs) by employing a paraphrasing strategy. The framework uses negative similarity, specifically a BERTScore metric, as a reward signal. By inputting prefixes from the original pre-training dataset into the LM to generate suffixes, the dissimilarity between the true and generated suffixes is calculated. This score is then maximized during training to ensure that the LM's tendency to replicate verbatim memorization is diminished.  Besides, they found that when model size increases, both the convergence rate and dissimilarity score increase, suggesting that larger models may tend to ``forget" the memorized data faster. 

\subsubsection{Machine Unlearning}\label{sec:llm4}
Even if the LLM has memorized some copyrighted data, the model builder can thereafter use ``machine unlearning'' techniques to delete the copyrighted information for copyright protection. 
For example, \citet{eldan2023s}  first used a reinforced model to identify key tokens by comparing logits with a baseline model. Then, unique expressions in the data are replaced with generic ones, and new labels are generated to mimic a model not trained on the data. Finally, they show that fine-tuning the model with these labels can effectively remove the original text from the model's memory upon contextual prompting.  \citet{chen2023unlearn}  proposed an efficient method EUL for updating Large Language Models (LLMs) without full retraining by integrating ``lightweight unlearning layers''. The 
\begin{wrapfigure}{r}{0.55\textwidth} 
    \centering
    \vspace{-0.18in}
    \includegraphics[width=\linewidth]{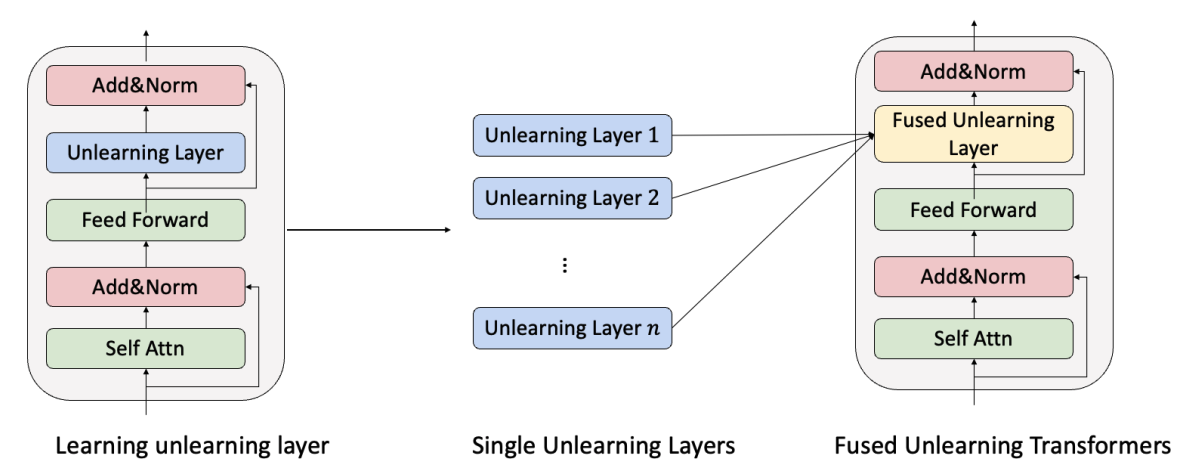}
    \vspace{-0.3in}
    \caption{\small  An overview of EUL pipeline (Image Credit to \citet{chen2023unlearn})}
    \label{fig:EUL}
    \vspace{-0.1in}
\end{wrapfigure}
pipeline of EUL is shown in Figure \ref{fig:EUL}. These layers are trained using a selective teacher-student objective within the transformer architecture. Additionally, to handle a sequence of forgetting operations, a fusion mechanism is employed to merge various unlearning layers efficiently where each learns to forget different sets of data. 
\citet{yao2023large} proposed a resource-efficient unlearning method that only  requires negative examples that we want the LLM to forget, without the need  for access to the original training set.  Their method mainly utilizes gradient ascent to make the model forget the undesired information, like user-reported or red-teaming failed cases. Besides, they also added additional loss terms to enforce the model generated similar outputs with the unlearned model on some normal inputs to maintain their utility.  \citet{pawelczyk2023context} explored a new class of unlearning methods for LLMs called “In-Context Unlearning”(ICUL) without access to the model parameters.  ICUL first inverts the label of the data to be forgotten and then incorporates a set of correctly labeled data points to mitigate over-correction, finally, it integrates the query input into the altered training template and prompts the model to predict with no temperature adjustment to erase the specified data from the model's knowledge. Their empirical results suggest that ICUL reliably removes the influence of training points and show that label flipping for in-context examples can impact the model’s output.

\subsection{Model Copyright Protection}\label{sec:llmb}

The development of LLMs such as ChatGPT/GPT-4 has significantly enhanced their ability to generate human-like texts. Many companies have integrated LLMs into interactive user interfaces like New Bing, yielding numerous benefits. However, the emergence of LLMs also brings forth new challenges. On one hand, LLM increased the difficulty in distinguishing between genuine content and fake news, propaganda, or misinformation generated by LLMs like GPT-4. On the other hand, LLMs are susceptible to theft. For example, ~\citet{alpaca} demonstrated the possibility of distilling LLMs solely through API access.  
This underscores the importance of safeguarding the copyright of LLMs, which often require significant resources for training. In light of these concerns, the protection methods can be classified into two categories based on their objectives:
\begin{itemize}[noitemsep,topsep=0pt]
    \item \textbf{Watermark}, which ascertains whether given texts are generated by the LLMs owned by the respective company. 
    \item \textbf{Model parameter protection}, which detects unauthorized model stealing the proprietary model parameters.
\end{itemize}

In the following of this section, we discuss the protection on text generation model from the two perspectives.


\subsubsection{Watermark} 
This type of protection aims to detect whether a text is generated by a particular LLM. Watermarking-based methods are majorly explored to detect LLM-generated texts. In general, this type of method attaches a watermark generator with LLMs which modify the generated texts to include pre-designed watermarks and a detector is used to determine whether suspicious texts contain watermarks. In this subsection, we conduct a comprehensive analysis of various watermarking methods, with a focus on discussing their robustness and the trade-offs between detection accuracy and text quality.

In \citet{kirchenbauer2023watermark}, LLM-generated texts with watermarks are generated by forcing the model to sample words from a partition of vocabulary with a higher probability. In detail, a ``green list'' is a random partition of the vocabulary decided by a random function with the hash of the previous token as the random seed. During the text generation process of an LLM, the probability of green list is increased in the sampling step. Thus, tokens from green list has a high probability of being sampled. Since the green list is decided by the previous tokens, this operation forms the matching between green lists and the consequent tokens. A higher proportion of token from green list indicate the text is more possible to generated by a watermarked model. A $z$-test is conducted to detect this matching between tokens and determine whether the test texts is generated with or without knowledge of the green list. The texts that have a significant $z$-score will be detected as LLM-generated texts. 
Based on the work~\citep{kirchenbauer2023watermark}, \citet{yoo2023advancing} improved it by partitioning the vocabulary into lists of more colors instead of just green list. This method can encode a message like the ownership information into the text, and enrich the application of watermarks. \citet{takezawa2023necessary} highlighted that longer generated texts typically exhibit higher $z$-scores. To accommodate this, it may be reasonable to relax the strict proportion requirements of the green list for lengthier texts. 
 
However, due to the fact that the watermark in the texts is easy to be perturbed by attack methods like paraphrasing, synonym replacement and so on, more works focusing on the robustness of watermarks are proposed. For example, in the watermarking method of \citet{kirchenbauer2023reliability}, once the previous token is reworded by paraphrase, the seed of the random function will also change which causes a different green list and disturb the matching between the green list and the consequent token. \citet{kirchenbauer2023reliability} leveraged an improved hashing scheme to enhance the reliability and robustness of this watermarking in real scenarios. This method uses more previous tokens to compute the random seed to increase the tolerance of paraphrase if part of the tokens are reworded, and a windowed test called WinMax is leveraged in the detection process. \citet{zhao2023provable} proposed Unigram-Watermark against paraphrasing attacks. It changes the method of \citet{kirchenbauer2023watermark} with a fixed ``green list'' which does not change no matter what the hash of the previous token is. They proved that the fixed list can also be secure and robust. However, this robustness requires a constraint that only a limited amount of words can be changed, which is strict in practical scenarios. 

\citet{ren2023robust} proposed the method called SemaMark to use the semantics as the seed of the hash function. Because the hash of the token is easy to change by paraphrase, but the semantics are usually stable. SemaMark uses the embeddings of the previous tokens, which represent the semantic information, as the seed of the random function to decide the green list. The embeddings can be obtained from the LLM itself. Even though the malicious user wants to use paraphrase to remove the watermark for copyright infringement, the semantics are likely to remain unchanged and will not influence the partition of the green list for the consequent token. Thus, the watermark is robust and will not be easily removed by paraphrase. ~\citet{liu2023semantic} proposed a similar method which uses an additional text embedding model to extract the embeddings.


\citet{fu2023watermarking} found that the random selection of green list might be conflicted with the original semantics. For example, the tokens in the generated output for a summarization task will have a overlap with the input text to summarize. Thus, \citet{fu2023watermarking} proposed to incorporate the semantically related tokens into green list and randomly partition the rest vocabulary. The semantically related tokens are selected by the similarity with the previous inputs. This can further increase the fraction of green list matching in watermarked text and thus improve the detection performance.


To safeguard against the deciphering of the green list's watermark rule, \citet{liu2023private} introduced an unfalsifiable approach. This technique employs a fully connected layer to ascertain the green list for each token position, taking into account preceding tokens. The input for this layer comprises a pre-established window of prior tokens. \citet{liu2023private} claimed that their method resists cracking owing to the substantial size of the window and the difficulty in extracting the network for the green list by external parties, primarily because of the absence of clear ground truth labels.


The previous watermarking methods rely on the ``green list'', which is shown to harm the quality of generated texts \citep{sato2023embarrassingly, takezawa2023necessary, hu2023unbiased, kuditipudi2023robust, christ2023undetectable}. \citet{kuditipudi2023robust} proposed a distortion-free watermarking scheme and took the randomness of watermark key $\xi$ into consideration. They assume the watermarked generation given a random watermark key $\xi_i$ is $\Gamma(\xi_i,p_F)$ where $p_F$ is defined in Eq. \eqref{eq:ps}. Then the watermark is claimed as distortion-free if 
\begin{equation}\label{eq:distort-free}
    \mathbb{P}(\Gamma(\xi_i,p_F)=w_i)=p_F(w_i)
\end{equation}

In particular, they implement the $\Gamma$ with a decoder that maps a sequence of uniform random variables and permutations to tokens using inverse transform sampling as follows:
$$
\Gamma(\xi, p_F):=\pi^{-1}(\min\{\pi(i):p_F(\{j:\pi(j)\le \pi(i)\})\ge u\})
$$
In this equation, the watermark key $\xi=(u,v)$, $\pi$ is a random permutation and $u$ is a pre-specified threshold. With this formulation, $\Gamma(\xi,p_F)$ is designed to select the token possessing the smallest index within the permutation $\pi$ ensuring that CDF of $p_F$ related to $\pi$ reaches a minimum threshold of $u$. For identifying watermarked text, the detection process involves correlating the token indices within the text against $\xi$. On the contrary, the indices of tokens in the non-watermarked text will be i.i.d uniform irrespective of the text itself and thus not correlated to the watermark key $\xi$. This scheme is shown to be distortion-free with respective to the definition in Eq.~\eqref{eq:distort-free}.

\citet{hu2023unbiased} also proposed a watermark called Unbiased Watermark aiming to guarantee the non-degradation of the text quality. This method leverages distribution reweighting to modify the original generation distribution into the watermarked distribution. In detail, given a watermark code $E$, they define a reweighting function $R_E$ mapping the distribution $p_F(\cdot|w_1,...,w_{i-1})$ to the watermarked version $R_E(p_F(\cdot|w_1,...,w_{i-1}))$. When taking the randomness of $E$ into consideration, the watermark is claimed as unbiased if and only if 
\begin{equation}\label{eq:unbiased water}
\mathbb{E}_{E}R_E(p_F(\cdot|w_1,...,w_{i-1}))=p_F(\cdot|w_1,...,w_{i-1})
\end{equation}
This equation indicates that although the generation distribution for each individual token is distorted, the mean output token probabilities remain the original distribution when considering the randomness of $E$.
In practice, they first select a 1024-bit random bitstring as a key $k$, then the watermark code $E$ is generated via a hash function $\hat{E}$ whose inputs are $k$ and the token sequence $(w_1,...,w_{i-1})$, i.e. $E=\hat{E}(k, (w_1,...,w_{i-1}))$. Finally, a delta function $R_E(P)=\delta_{sampling_P(E)}$ is used as the reweighting function to create the watermarked distribution. They showed that this procedure satisfies Eq.~\eqref{eq:unbiased water}, thus unbiased. During the detection stage, they leverage the likelihood ratio test and compute a log likelihood ratio (LLR) score to detect the watermarks. 

The above watermark methods require the access to generation process and cannot be used to the scenarios of black-box models or when there is only API for the LLM available. \citet{yang2023watermarking} developed a binary-encoding-based watermarking framework for black-box language models. To be more specific, original words in generated texts are encoded as bit-0, and part of their synonym candidates are encoded as bit-1. Then a hash function is leveraged to randomly substitute bit-0 words with bit-1 synonyms, and a $z$-test is conducted to infer the change of distribution for bit-1 and bit-0.

From a different perspective, a family of watermarks named Easymark is proposed in \citep{sato2023embarrassingly} to add watermarks directly to the generated texts rather than the generation distribution. Whitemark, the most representative method in this family, exploits the fact that Unicode has many codepoints for whitespace and it replaces a whitespace from the original codepoint U+0020 to another codepoint such as U+2004. Then during the detection process, the number of U+2004 is counted to determine if the text is watermarked. Therefore, this method does not distort the distribution during generation and does not change the original meaning of the text. 
 
\subsubsection{Model Parameter Protection} 
\begin{wrapfigure}{r}{0.35\textwidth} 
    \centering
    \vspace{-0.18in}
    \includegraphics[width=\linewidth]{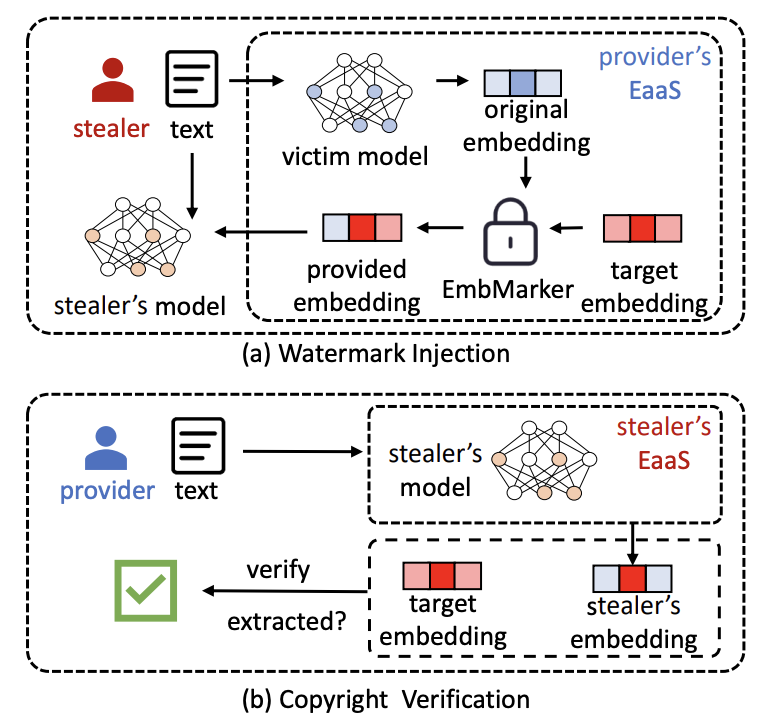}
    \vspace{-0.3in}
    \caption{\small  An overview of EmbMarker (\citet{peng2023you})}
    \label{fig:EmbMarker}
    \vspace{-0.1in}
\end{wrapfigure}
This type of protection aims to detect unauthorized models stealing the proprietary model parameters. \citet{birch2023model} proposed the Model Leeching attack method, which can distill characteristics by questioning an LLM. This method can extract task capability from ChatGPT-3.5-Turbo, achieving 73\% Exact Match (EM) similarity and SQuAD EM and F1 accuracy scores of 75\% and 87\% with only \$50 cost \citet{birch2023model}. After getting an extracted model, attackers can then inverse the model, inference membership, leak privacy data, and theft model intellectual property. \citet{peng2023you} proposed an Embedding Watermark method called EmbMarker which can defend against model extraction attacks from Embedding as a Service (EaaS). As shown in Figure~\ref{fig:EmbMarker}, EmbMarker selects a set of moderately frequent words from a general text corpus to create a trigger set. It then chooses a target embedding to serve as the watermark. This watermark is inserted into the embeddings of texts that contain trigger words, functioning as a backdoor. The intensity of the watermark insertion is proportional to the quantity of trigger words present in the text. This method ensures the efficient transfer of the watermark backdoor to the EaaS-stealer's model for copyright verification purposes, while simultaneously minimizing any negative effects on the original utility of the embeddings. \citet{fan2023fate} proposed FateLLM, an industrial-grade federated learning framework for LLMs, which can protect the intellectual property of LLMs by using a federated intellectual property protection approach. Every client autonomously confirms the presence of watermarks in the model, asserting their individual ownership of the federated model. This verification is achieved without revealing any private training data or confidential watermark details. Additionally, the approach ensures 
the preservation of data privacy throughout both the training and inference phases by employing privacy-protective mechanisms. \citet{he2022cater} uncovered that prevailing watermarking techniques often disrupt the word distribution. This distortion can be exploited to deduce the watermarked words by analyzing the frequency changes in potential watermark candidates through sufficient statistical methods. Consequently, this makes the watermarks relatively easy to identify and remove. Therefore, they proposed CATER for protecting text generation APIs via more stealthy watermarks. In detail, CATER injects the watermarks in a word distribution conditional on linguistic features (condition $c$) such as part-of-speech and dependency tree \citep{nadkarni2011natural}, while maintaining the original word distribution. 


\section{Other domains}\label{sec:others}

In this section, we will discuss copyright protection in various domains, including code and audio. It is important to acknowledge that, in contrast to areas like LLMs and image generation, the fields of code and audio generation have relatively fewer studies focused on copyright protection against DGMs. We will outline and discuss the existing methods employed in the generation of code and audio in this segment.


\subsection{Code Generation}


Code copyright is a critical aspect of the text domain with distinctive characteristics. As opposed to other forms of textual content, codes are often accompanied by explicit copyright licenses, and code owners are able to take more proactive strategies to protect their data.
In this survey, code copyright protection majorly refers to preventing the usage of copyrighted code as a part of the training data for machine learning models without the explicit permission of the copyright holder. For example, Copilot (\href{https://copilot.github.com/}{https://copilot.github.com/}), a closed-source deep learning code generation model, is trained using a wide range of open-source code repositories sourced from GitHub, without consideration for their licenses. It has been observed that Copilot occasionally reproduces identical code snippets with copyleft licenses from its training dataset while generating code~\citep{lee2023wrote}.
As these models learn from the data they are trained on, they can generate new code that is strikingly similar to the copyrighted code, potentially leading to copyright infringement, which poses a significant threat to code copyright protection.

There are different aspects to protect code copyright. From the model builder's perspective, they can carefully curate their training data and train the model to provide reference during generation, so that the data sources are legal and can be tracked. For example, \citet{li2023starcoder} took several important steps to ensure safe release of their code-LLMs StarCoder and  StarCoderBase, including an improved Personally Identifiable Information~(PII) redaction pipeline and a novel attribution tracing tool, making the StarCoder models publicly available under a more commercially viable version of the Open Responsible AI Model license. To remove PII from the training data, the authors trained an encoder-only model called StarEncoder and then fine-tuned it for the Named Entity Recognition (NER) task to detect PII. \citet{yan2022whygen} introduced a tool named WhyGen that explains the generated code by referring to training examples. The method extracts an inference fingerprint from the neural network when generating the code and using the inference fingerprint to get the most similar examples from training data. The inference fingerprint is produced by the activation values of a set of intermediate neurons in the network during the inference pass. The authors claimed that if Whygen with the auto-completion feature is used in an IDE, it can help reduce the concern about using source code from copyrighted project without permission.

From the perspective of code owners, they have the solutions to proactively protect their codes via poisoning or watermarking. \citet{lee2023wrote} designed and implemented a prototype, CoProtector, which uses data poisoning techniques to defend source code repositories against such exploits. Specifically, given a repository to protect, CoProtector generates two distinct types of poison instances from its original code artifact. One type aims to reduce the performance of Copilot-like DGMs, and the other can reliably embed backdoors for later detection. In addition, CoProtector proposes a collaborative protection framework that manages protected poison repositories and amount of poison instances in the network. \citet{ji2022unlearnable} designed a set of lightweight transformations that can be applied to codes before they are open-source released. They also leveraged a pre-trained code embedding model (e.g., CodeBERT~\citep{feng2020codebert}) to guide the selection of transformations. The transformed codes are called ``unlearnable examples'' so that code generation models trained on these codes can only obtain faulty knowledge. 

Besides, code owners can also check the safety of their source codes via membership inference. \citet{ma2023code} proposed CodeForensic, which uses membership inference to audit whether codes are misused for training a code generator. Particularly, it leverages a reference-model-based likelihood ratio test (LRT) to ascertain if a specific code snippet is present in a model's training dataset. The experimental outcomes indicate that the LRT-based membership inference can deliver satisfactory results. However, the method is limited in practical application due to its reliance on the access to logits of the generative model.

\subsection{Audio Generation}
Utilizing zero-shot voice synthesis technology, it is now possible to generate realistic audio from just a few seconds of a speaker's recording. Consequently, the implementation of watermarking in the audio domain is essential to prevent model misuse and data infringement. Among existing methods, the primary focus is on protection of model copyright.

\citet{Chen2023WavMark} introduced a post-watermarking method called WavMark to accelerate the detection synthesized audio and prevent the unauthorized usage of audio generation models. The post-watermarking process, as applied in WavMark, modifies the audio after its generation. It incorporates a watermark by encoding a message into the frequency domain of the generated audio. The model owner can detect the watermark by extracting the message from suspect audios. The whole framework of WavMark is a invertible neural network that is composed of $L$ cascaded invertible blocks. Each block can be applied in both a encoding way and a decoding way. WavMark first transforms the audio $\mathbf{x}$ to watermark into frequency domain by a short-time Fourier, which is denoted as $\mathbf{x}_{\text{frq}}$. Then the network encodes a 32-bit message $m$ into the audio $\mathbf{x}_{\text{frq}}$ by the encoding operation in each block:
\begin{align*}
\mathbf{x}_{\text{frq}}^{l+1} =\mathbf{x}_{\text{frq}}^l+\phi\left(\mathbf{m}^l\right), \quad \text{and} \quad
\mathbf{m}^{l+1} =\mathbf{m}^l \odot \exp \left(\sigma\left(\rho\left(\mathbf{x}_{\text{frq}}^{l+1}\right)\right)\right)+\eta\left(\mathbf{x}_{\text{frq}}^{l+1}\right),
\end{align*}
where $\phi$, $\rho$, and $\eta$ is a dense block introduced by~\citet{wang2018esrgan}, and $\mathbf{m}^l$ and $\mathbf{x}_{\text{frq}}^{l}$ is the message and audio data in $l$-th block. The decoding operation is starting from $\mathbf{x}_{\text{frq}}^{L}$ and a random sampled starting point of message $\mathbf{m}^L$:
\begin{align*}
\mathbf{m}^l =\left(\mathbf{m}^{l+1}-\eta\left(\mathbf{x}^{l+1}\right)\right) \odot \exp \left(-\sigma\left(\rho\left(\mathbf{x}^{l+1}\right)\right)\right) \quad \text{and} \quad 
\mathbf{x}^l =\mathbf{x}^{l+1}-\phi\left(\mathbf{m}^l\right)
\end{align*}
The invertible neural network is trained to minimized the $L_2$ distance of encoded and decoded messages. In this way, WavMark can encode and decode a message from the frequency domain of an generated audio.


\citet{Cao2023MelspectMark} proposed a watermarking method to protect the copyright of audio-generative diffusion models. This method follows the main idea of trigger-based watermarking in Section~\ref{sec:image_trigger_based}. It incorporates a trigger mechanism into the diffusion model. Once the trigger is activated by a pre-defined input, the model will generate a watermarked audio to verify the ownership of the audio-generative diffusion model. To make the trigger invisible, this method uses environmental natural sounds and Infrasounds around 10Hz, which are typically ignored in daily life. From an acoustic standpoint, environmental sounds blend with ambient noise encountered by machines, while Infrasounds are inaudible to humans and indiscernible in mel-spectrograms. Additionally, it’s important that these watermarking triggers do not significantly impact the performance of the original diffusion model, so that the generated samples remain true to the training data if there is no trigger in the input of the diffusion model.

\section{Discussion}

While a range of technical approaches have been suggested, numerous challenges remain unresolved in this field. In this section we will explore the existing issues and consider potential future directions for copyright protection within deep generative models. 

\textbf{First}, we discuss the data copyright protection from the following three perspectives:

$\bullet$ \textit{Comprehensiveness}. In image generation, most of the protection for data copyright conducted by the source data owner is tailored for a specific model or learning algorithm. However, they cannot prevent the model builder from learning the data with a different generative model, which means they cannot provide a comprehensive protection against different DGMs. For example, the unrecognizable examples by~\citet{shan2023glaze, liang2023mist} focus on Stable Diffusion, and~\citet{van2023anti} prevent the fine-tuning scheme of DreamBooth. The watermark like methods proposed by~\citet{ma2023generative, wang2023detect} is also designed for one type of DGMs. 
A dive to the effectiveness of these methods in safeguarding data against a range of generative models, rather than just  tailored ones, is essential. 
A thorough protection encompassing a wide array of DGMs is significantly more meaningful.
Its challenge stems from the distinct architectures and unique properties of various generative models, making it challenging to devise protections for specific vulnerabilities. Despite these obstacles, pursuing comprehensive research in this field is essential and may lead to a deeper understanding of DGM characteristics.


$\bullet$ \textit{Data protection by owners besides image domain.} Although several methods to protect the data copyright are developed to the source data owners in image domain~\cite{shan2023glaze, liang2023mist, cui2023diffusionshield, webster2023duplication}, there are fewer works for text and other domains. On the one hand, the DGMs in other domains like the code, audio, graph and video do not have the same level of development as image. On the other hand, the protection in some domains like text is much harder than image, because the modification on image can be designed as invisible to human eyes but effective on DGMs, while text is discrete and is hard to be designed as imperceptible. Despite the difficulty, it is necessary to propose the protection from the side of source data owner for other domains, especially in text domain which is fast developing and causes increasing concerns in the data copyright. Also, the copyright protection for multi-modality generation~\cite{ma2023unified, zhan2023multimodal, ruan2023mm, pichler2010pet} in both data copyright and model copyright is also crucial.


$\bullet$ \textit{Infringement detection.} Injecting watermark into images to accelerate the infringement detection has been introduced in Section~\ref{sec:image_watermark_source_data_owner}, which aims to help source data owners protect their data copyright. Infringement detection can also benefit the model builder. Before releasing the generated output to the users, the builder can first check whether the output is infringing the copyright of training data (like a memorized training sample). If the output to be released is detected as infringement, the model builder can use some following-up strategies like adding references to the source of the content or removing the suspicious part to avoid the infringement. However, detecting infringement on the training data is not trivial, especially that current DGMs usually require a large amount of training data to ensure the generated quality and diversity. The dataset de-duplication in Section~\ref{sec:data_duplication} and Section~\ref{sec:llm1} provides potential solutions. It is an offline strategy that can be achieved before training. However, for infringement detection, a faster real-time searching is necessary to mitigate negative impact on the generation speed. 
Another issue is that, compared with memorization problem, the infringement on the copyright of abstract concepts, such as the style of an artwork or a product, and the original storylines and characters created by authors, is hard to confirm. The memorized samples often bear a high resemblance to the copyrighted material\footnote{\href{https://spectrum.ieee.org/midjourney-copyright}{https://spectrum.ieee.org/midjourney-copyright}, \href{https://www.nytimes.com/2023/12/27/business/media/new-york-times-open-ai-microsoft-lawsuit.html}{https://www.nytimes.com/2023/12/27/business/media/new-york-times-open-ai-microsoft-lawsuit.html}}, making infringement easy to confirm. In contrast, if a generated sample replicates the style of a picture but diverges significantly in content from the original, it becomes much harder to detect and confirm whether the generated sample stems from copyrighted data or not.





\textbf{Second}, for the model copyright protection, we present the following aspects:

$\bullet$ \textit{Robustness.} The purification techniques for watermark and adversarial perturbations have challenged the robustness of watermark protection on model copyright in both image generation~\cite{nie2022diffusion, yoon2021adversarial, chen2021refit} and text generation~\cite{krishna2023paraphrasing, ren2023robust}. In image generation models, the imperceptibility of watermarks is a key reason why the watermark is easy to remove by methods like denoising. In text generation models, the watermark is vulnerable because the watermark is encoded based on the tokens, and thus operations such as paraphrasing can easily change the tokens by rewording and reordering. In fact, the robustness issue also exists in the protection on data copyright like unrecognisable examples and watermarks for source data owners.


$\bullet$ \textit{Trade-off between protection and performance.} The introduction of watermarks can potentially reduce the generation quality. \citet{wen2023treering} and ~\citet{cui2023ft} found that the watermarks will cause influence on the generated contents by Stable Diffusion. The trajectory of reverse process in diffusion model is sensitive to the change since the change might be accumulated in the long generation steps. \citet{ajith2023performance} also pointed out that the watermark method on LLMs will reduce the generated quality on the long-term generation task like summarization. Thus, developing protection methods that can maintain the performance of the model is necessary.


$\bullet$ \textit{Flexibility.} 
The protection is usually designed for one specific type of models and lack of flexibility. For example, in Section~\ref{sec:model_protection}, the methods proposed by~\cite{ong2021protecting}, \cite{zhang2023generative}, \cite{fei2022supervised}, \cite{yu2020responsible} and \cite{yu2021artificial} are all tailored for GANs. The effectiveness of the protection on other model architectures are not guaranteed. 
If one protection method is adaptable to a variety of DGMs, it eliminates the need for redundant efforts in redesigning for different models, thereby enhancing the efficiency of the protection process. This opens up opportunities for third-party providers to offer such protective services. They could utilize a versatile method to secure different DGMs without the requirement for constant redesigning. This strategy not only streamlines the process but also represents a vital direction in the realm of model copyright protection.


\textbf{For the epilogue}, we recognize that the issue of copyright protection in deep generative models remains a work in progress, encompassing both data and model security. The rapid evolution of DGMs is catalyzing the establishment of effective regulatory oversight, fostering healthy market dynamics, and facilitating the empowerment of various industries. This progression demands not only improvements in model performance but also a critical evaluation of how these technologies are utilized, ensuring their use is reasonable and standardized. Copyright protection stands as a pivotal concern, and it is anticipated that in the foreseeable future, it will garner increased attention and resources from both industrial and academic spheres. Our hope is that this work can serve as a guiding framework and a source of inspiration for possible forthcoming directions in the realm of copyright protection for future researchers who wants to develop tools in this field from the perspective of techniques.

\bibliographystyle{ACM-Reference-Format}
\bibliography{sample-base}










\end{document}